\documentclass[longauth]{aa}  
\let\ACMmaketitle=\maketitle
\renewcommand{\maketitle}{\begingroup\let\footnote=\thanks \ACMmaketitle\endgroup}
\usepackage{euclid}
\usepackage{amssymb}
\usepackage{graphicx}
\usepackage{nicefrac}
\usepackage{amsmath}
\usepackage{booktabs}
\usepackage{siunitx}
\usepackage{diagbox}
\usepackage{float}
\usepackage{gensymb}
\usepackage{txfonts}
\usepackage{natbib}

\makeatletter
\providecommand\add@text{}
\newcommand\tagaddtext[1]{%
  \gdef\add@text{#1\gdef\add@text{}}}%
\renewcommand\tagform@[1]{%
  \maketag@@@{\llap{\add@text\quad}(\ignorespaces#1\unskip\@@italiccorr)}%
}
\newcommand{\magarc}{mag arcsec\ensuremath{^{\mathrm{-2}}}}
\newcommand{\mulim}{\ensuremath{\mu_{\rm lim}}}
\newcommand{\muvis}{\ensuremath{\mu_{\scriptscriptstyle\rm VIS,AB}}}
\newcommand{\muzody}{\ensuremath{\mu_{\rm zodi}}}
\newcommand{\mucib}{\ensuremath{\mu_{\scriptscriptstyle\rm CIB}}}
\newcommand{\mustray}{\ensuremath{\mu_{\rm stray}}}
\newcommand{\muism}{\ensuremath{\mu_{\scriptscriptstyle\rm ISM}}}

\usepackage{hyperref}
\hypersetup{colorlinks,linkcolor=black,citecolor=blue,filecolor=cyan,urlcolor=magenta}

\begin{document}

\title{\Euclid preparation: XVI. Exploring the ultra low-surface brightness Universe with Euclid/VIS}
\titlerunning{Exploring the ultra-low surface brightness Universe with \Euclid/VIS}

\author{Euclid Collaboration: A.S.~ Borlaff$^{1,2}$\thanks{\email{a.s.borlaff@nasa.gov}}, P.~G\'omez-Alvarez$^{2,3}$, B.~Altieri$^{2}$, P.M.~Marcum$^{1}$, R.~Vavrek$^{2}$, R.~Laureijs$^{4}$, R.~Kohley$^{2}$, F.~Buitrago$^{5,6}$, J.-C.~Cuillandre$^{7,8}$, P.-A.~Duc$^{9}$
% ...
}

\authorrunning{Borlaff et al.}

\institute{$^{1}$ NASA Ames Research Center, Moffett Field, CA 94035, USA\\
$^{2}$ ESAC/ESA, Camino Bajo del Castillo, s/n., Urb. Villafranca del Castillo, 28692 Villanueva de la Ca\~nada, Madrid, Spain\\
$^{3}$ FRACTAL S.L.N.E., calle Tulip\'an 2, Portal 13 1A, 28231, Las Rozas de Madrid, Spain\\
$^{4}$ European Space Agency/ESTEC, Keplerlaan 1, 2201 AZ Noordwijk, The Netherlands\\
$^{5}$ Departamento de F\'{i}sica Te\'{o}rica, At\'{o}mica y \'{O}ptica, Universidad de Valladolid, 47011 Valladolid, Spain\\
$^{6}$ Instituto de Astrof\'isica e Ci\^encias do Espa\c{c}o, Faculdade de Ci\^encias, Universidade de Lisboa, Tapada da Ajuda, PT-1349-018 Lisboa, Portugal\\
$^{7}$ AIM, CEA, CNRS, Universit\'{e} Paris-Saclay, Universit\'{e} de Paris, F-91191 Gif-sur-Yvette, France\\
$^{8}$ Observatoire de Paris, PSL Research University 61, avenue de l'Observatoire, F-75014 Paris, France\\
$^{9}$ Observatoire Astronomique de Strasbourg (ObAS), Universit\'e de Strasbourg - CNRS, UMR 7550, Strasbourg, France\\
$^{10}$ Institute of Cosmology and Gravitation, University of Portsmouth, Portsmouth PO1 3FX, UK\\
$^{11}$ INAF-Osservatorio Astronomico di Brera, Via Brera 28, I-20122 Milano, Italy\\
$^{12}$ INAF-Osservatorio di Astrofisica e Scienza dello Spazio di Bologna, Via Piero Gobetti 93/3, I-40129 Bologna, Italy\\
$^{13}$ Mullard Space Science Laboratory, University College London, Holmbury St Mary, Dorking, Surrey RH5 6NT, UK\\
$^{14}$ IFPU, Institute for Fundamental Physics of the Universe, via Beirut 2, 34151 Trieste, Italy\\
$^{15}$ SISSA, International School for Advanced Studies, Via Bonomea 265, I-34136 Trieste TS, Italy\\
$^{16}$ INFN, Sezione di Trieste, Via Valerio 2, I-34127 Trieste TS, Italy\\
$^{17}$ INAF-Osservatorio Astronomico di Trieste, Via G. B. Tiepolo 11, I-34131 Trieste, Italy\\
$^{18}$ Universidad de la Laguna, E-38206, San Crist\'{o}bal de La Laguna, Tenerife, Spain\\
$^{19}$ Instituto de Astrof\'{i}sica de Canarias, Calle V\'{i}a L\`actea s/n, 38204, San Crist\`obal de la Laguna, Tenerife, Spain\\
$^{20}$ Dipartimento di Fisica e Astronomia, Universit\'a di Bologna, Via Gobetti 93/2, I-40129 Bologna, Italy\\
$^{21}$ INFN-Sezione di Bologna, Viale Berti Pichat 6/2, I-40127 Bologna, Italy\\
$^{22}$ Universit\"ats-Sternwarte M\"unchen, Fakult\"at f\"ur Physik, Ludwig-Maximilians-Universit\"at M\"unchen, Scheinerstrasse 1, 81679 M\"unchen, Germany\\
$^{23}$ Max Planck Institute for Extraterrestrial Physics, Giessenbachstr. 1, D-85748 Garching, Germany\\
$^{24}$ INAF-Osservatorio Astrofisico di Torino, Via Osservatorio 20, I-10025 Pino Torinese (TO), Italy\\
$^{25}$ Department of Astronomy, University of Geneva, ch. d\'Ecogia 16, CH-1290 Versoix, Switzerland\\
$^{26}$ INFN-Sezione di Roma Tre, Via della Vasca Navale 84, I-00146, Roma, Italy\\
$^{27}$ Department of Mathematics and Physics, Roma Tre University, Via della Vasca Navale 84, I-00146 Rome, Italy\\
$^{28}$ INAF-Osservatorio Astronomico di Roma, Via Frascati 33, I-00078 Monteporzio Catone, Italy\\
$^{29}$ INAF-Osservatorio Astronomico di Capodimonte, Via Moiariello 16, I-80131 Napoli, Italy\\
$^{30}$ Centro de Astrof\'{\i}sica da Universidade do Porto, Rua das Estrelas, 4150-762 Porto, Portugal\\
$^{31}$ Instituto de Astrof\'isica e Ci\^encias do Espa\c{c}o, Universidade do Porto, CAUP, Rua das Estrelas, PT4150-762 Porto, Portugal\\
$^{32}$ INFN-Bologna, Via Irnerio 46, I-40126 Bologna, Italy\\
$^{33}$ Dipartimento di Fisica e Scienze della Terra, Universit\'a degli Studi di Ferrara, Via Giuseppe Saragat 1, I-44122 Ferrara, Italy\\
$^{34}$ INAF, Istituto di Radioastronomia, Via Piero Gobetti 101, I-40129 Bologna, Italy\\
$^{35}$ Institut de Recherche en Astrophysique et Plan\'etologie (IRAP), Universit\'e de Toulouse, CNRS, UPS, CNES, 14 Av. Edouard Belin, F-31400 Toulouse, France\\
$^{36}$ INFN-Sezione di Torino, Via P. Giuria 1, I-10125 Torino, Italy\\
$^{37}$ Dipartimento di Fisica, Universit\'a degli Studi di Torino, Via P. Giuria 1, I-10125 Torino, Italy\\
$^{38}$ Universit\'e C\^{o}te d'Azur, Observatoire de la C\^{o}te d'Azur, CNRS, Laboratoire Lagrange, Bd de l'Observatoire, CS 34229, 06304 Nice cedex 4, France\\
$^{39}$ INAF-IASF Milano, Via Alfonso Corti 12, I-20133 Milano, Italy\\
$^{40}$ Institut de F\'{i}sica d’Altes Energies (IFAE), The Barcelona Institute of Science and Technology, Campus UAB, 08193 Bellaterra (Barcelona), Spain\\
$^{41}$ Institute of Space Sciences (ICE, CSIC), Campus UAB, Carrer de Can Magrans, s/n, 08193 Barcelona, Spain\\
$^{42}$ Institut d’Estudis Espacials de Catalunya (IEEC), Carrer Gran Capit\'a 2-4, 08034 Barcelona, Spain\\
$^{43}$ Dipartimento di Fisica e Astronomia “Augusto Righi” - Alma Mater Studiorum Università di Bologna, via Piero Gobetti 93/2, I-40129 Bologna, Italy\\
$^{44}$ Observatoire de Sauverny, Ecole Polytechnique F\'ed\'erale de Lau- sanne, CH-1290 Versoix, Switzerland\\
$^{45}$ Department of Physics "E. Pancini", University Federico II, Via Cinthia 6, I-80126, Napoli, Italy\\
$^{46}$ INFN section of Naples, Via Cinthia 6, I-80126, Napoli, Italy\\
$^{47}$ INAF-Osservatorio Astrofisico di Arcetri, Largo E. Fermi 5, I-50125, Firenze, Italy\\
$^{48}$ Institut national de physique nucl\'eaire et de physique des particules, 3 rue Michel-Ange, 75794 Paris C\'edex 16, France\\
$^{49}$ Centre National d'Etudes Spatiales, Toulouse, France\\
$^{50}$ Institute for Astronomy, University of Edinburgh, Royal Observatory, Blackford Hill, Edinburgh EH9 3HJ, UK\\
$^{51}$ Jodrell Bank Centre for Astrophysics, School of Physics and Astronomy, University of Manchester, Oxford Road, Manchester M13 9PL, UK\\
$^{52}$ European Space Agency/ESRIN, Largo Galileo Galilei 1, 00044 Frascati, Roma, Italy\\
$^{53}$ Univ Lyon, Univ Claude Bernard Lyon 1, CNRS/IN2P3, IP2I Lyon, UMR 5822, F-69622, Villeurbanne, France\\
$^{54}$ University of Lyon, UCB Lyon 1, CNRS/IN2P3, IUF, IP2I Lyon, France\\
$^{55}$ Departamento de F\'isica, Faculdade de Ci\^encias, Universidade de Lisboa, Edif\'icio C8, Campo Grande, PT1749-016 Lisboa, Portugal\\
$^{56}$ Instituto de Astrof\'isica e Ci\^encias do Espa\c{c}o, Faculdade de Ci\^encias, Universidade de Lisboa, Campo Grande, PT-1749-016 Lisboa, Portugal\\
$^{57}$ Universit\'e Paris-Saclay, CNRS, Institut d'astrophysique spatiale, 91405, Orsay, France\\
$^{58}$ Department of Physics, Oxford University, Keble Road, Oxford OX1 3RH, UK\\
$^{59}$ INFN-Padova, Via Marzolo 8, I-35131 Padova, Italy\\
$^{60}$ INAF-Istituto di Astrofisica e Planetologia Spaziali, via del Fosso del Cavaliere, 100, I-00100 Roma, Italy\\
$^{61}$ INAF-IASF Bologna, Via Piero Gobetti 101, I-40129 Bologna, Italy\\
$^{62}$ Institut d'Astrophysique de Paris, 98bis Boulevard Arago, F-75014, Paris, France\\
$^{63}$ Universit\'e de Paris, CNRS, Astroparticule et Cosmologie, F-75013 Paris, France\\
$^{64}$ Research Program in Systems Oncology, Faculty of Medicine, University of Helsinki, Helsinki, Finland\\
$^{65}$ Department of Physics, P.O. Box 64, 00014 University of Helsinki, Finland\\
$^{66}$ INAF-Osservatorio Astronomico di Padova, Via dell'Osservatorio 5, I-35122 Padova, Italy\\
$^{67}$ Institute of Theoretical Astrophysics, University of Oslo, P.O. Box 1029 Blindern, N-0315 Oslo, Norway\\
$^{68}$ Jet Propulsion Laboratory, California Institute of Technology, 4800 Oak Grove Drive, Pasadena, CA, 91109, USA\\
$^{69}$ von Hoerner \& Sulger GmbH, Schlo{\ss}Platz 8, D-68723 Schwetzingen, Germany\\
$^{70}$ Max-Planck-Institut f\"ur Astronomie, K\"onigstuhl 17, D-69117 Heidelberg, Germany\\
$^{71}$ Department of Physics and Helsinki Institute of Physics, Gustaf H\"allstr\"omin katu 2, 00014 University of Helsinki, Finland\\
$^{72}$ Aix-Marseille Univ, CNRS/IN2P3, CPPM, Marseille, France\\
$^{73}$ Universit\'e de Gen\`eve, D\'epartement de Physique Th\'eorique and Centre for Astroparticle Physics, 24 quai Ernest-Ansermet, CH-1211 Gen\`eve 4, Switzerland\\
$^{74}$ Leiden Observatory, Leiden University, Niels Bohrweg 2, 2333 CA Leiden, The Netherlands\\
$^{75}$ Helsinki Institute of Physics, Gustaf H{\"a}llstr{\"o}min katu 2, University of Helsinki, Helsinki, Finland\\
$^{76}$ NOVA optical infrared instrumentation group at ASTRON, Oude Hoogeveensedijk 4, 7991PD, Dwingeloo, The Netherlands\\
$^{77}$ Centre de Calcul de l'IN2P3, 21 avenue Pierre de Coubertin F-69627 Villeurbanne Cedex, France\\
$^{78}$ Dipartimento di Fisica "Aldo Pontremoli", Universit\'a degli Studi di Milano, Via Celoria 16, I-20133 Milano, Italy\\
$^{79}$ INFN-Sezione di Milano, Via Celoria 16, I-20133 Milano, Italy\\
$^{80}$ Argelander-Institut f\"ur Astronomie, Universit\"at Bonn, Auf dem H\"ugel 71, 53121 Bonn, Germany\\
$^{81}$ Instituto de F\'isica Te\'orica UAM-CSIC, Campus de Cantoblanco, E-28049 Madrid, Spain\\
$^{82}$ Aix-Marseille Univ, CNRS, CNES, LAM, Marseille, France\\
$^{83}$ Instituto de Astrof\'isica de Andaluc\'ia, CSIC, Glorieta de la Astronom\'\i a, E-18080, Granada, Spain\\
$^{84}$ Centre for Extragalactic Astronomy, Department of Physics, Durham University, South Road, Durham, DH1 3LE, UK\\
$^{85}$ Institut f\"ur Theoretische Physik, University of Heidelberg, Philosophenweg 16, 69120 Heidelberg, Germany\\
$^{86}$ Zentrum f\"ur Astronomie, Universit\"at Heidelberg, Philosophenweg 12, D- 69120 Heidelberg, Germany\\
$^{87}$ Istituto Nazionale di Astrofisica (INAF) - Osservatorio di Astrofisica e Scienza dello Spazio (OAS), Via Gobetti 93/3, I-40127 Bologna, Italy\\
$^{88}$ California institute of Technology, 1200 E California Blvd, Pasadena, CA 91125, USA\\
$^{89}$ ICC\&CEA, Department of Physics, Durham University, South Road, DH1 3LE, UK\\
$^{90}$ INFN, Sezione di Lecce, Via per Arnesano, CP-193, I-73100, Lecce, Italy\\
$^{91}$ Department of Mathematics and Physics E. De Giorgi, University of Salento, Via per Arnesano, CP-I93, I-73100, Lecce, Italy\\
$^{92}$ Department of Physics and Astronomy, University of Aarhus, Ny Munkegade 120, DK–8000 Aarhus C, Denmark\\
$^{93}$ Perimeter Institute for Theoretical Physics, Waterloo, Ontario N2L 2Y5, Canada\\
$^{94}$ Department of Physics and Astronomy, University of Waterloo, Waterloo, Ontario N2L 3G1, Canada\\
$^{95}$ Centre for Astrophysics, University of Waterloo, Waterloo, Ontario N2L 3G1, Canada\\
$^{96}$ Institute of Space Science, Bucharest, Ro-077125, Romania\\
$^{97}$ Institute for Computational Science, University of Zurich, Winterthurerstrasse 190, 8057 Zurich, Switzerland\\
$^{98}$ Departamento de Astrof\'{i}sica, Universidad de La Laguna, E-38206, La Laguna, Tenerife, Spain\\
$^{99}$ Dipartimento di Fisica e Astronomia “G.Galilei", Universit\'a di Padova, Via Marzolo 8, I-35131 Padova, Italy\\
$^{100}$ Departamento de F\'isica, FCFM, Universidad de Chile, Blanco Encalada 2008, Santiago, Chile\\
$^{101}$ Centre for Electronic Imaging, Open University, Walton Hall, Milton Keynes, MK7~6AA, UK\\
$^{102}$ Centro de Investigaciones Energ\'eticas, Medioambientales y Tecnol\'ogicas (CIEMAT), Avenida Complutense 40, 28040 Madrid, Spain\\
$^{103}$ Universidad Polit\'ecnica de Cartagena, Departamento de Electr\'onica y Tecnolog\'ia de Computadoras, 30202 Cartagena, Spain\\
$^{104}$ Kapteyn Astronomical Institute, University of Groningen, PO Box 800, 9700 AV Groningen, The Netherlands\\
$^{105}$ Department of Physics, P.O.Box 35 (YFL), 40014 University of Jyv\"askyl\"a, Finland\\
$^{106}$ Infrared Processing and Analysis Center, California Institute of Technology, Pasadena, CA 91125, USA\\
$^{107}$ Department of Physics and Astronomy, University College London, Gower Street, London WC1E 6BT, UK\\
}

\date{}

 
\abstract
{While \Euclid is an ESA mission specifically designed to investigate the nature of Dark Energy and Dark Matter, the planned unprecedented combination of survey area ($\sim15\,000$ deg$^2$), spatial resolution, low sky-background, and depth also make \Euclid an excellent space observatory for the study of the low surface brightness Universe. Scientific exploitation of the extended low surface brightness structures requires dedicated calibration procedures yet to be tested.}
{We investigate the capabilities of \Euclid to detect extended low surface brightness structure by identifying and quantifying sky background sources and stray-light contamination. We test the feasibility of generating sky flat-fields to reduce large-scale residual gradients in order to reveal the extended emission of galaxies observed in the \Euclid Survey.}
{We simulate a realistic set of \Euclid/VIS observations, taking into account both instrumental and astronomical sources of contamination, including cosmic rays, stray-light, zodiacal light, ISM, and the CIB, while simulating the effects of the presence of background sources in the FOV.}
{We demonstrate that a combination of calibration lamps, sky flats and self-calibration would enable recovery of emission at a limiting surface brightness magnitude of $\mulim=29.5^{+0.08}_{-0.27} $ \magarc\ ($3\sigma$, $10\times10$ arcsec$^2$) in the Wide Survey, reaching regions 2 magnitudes deeper in the Deep Surveys.}
{\Euclid/VIS has the potential to be an excellent low surface brightness observatory. Covering the gap between pixel-to-pixel calibration lamp flats and self-calibration observations for large scales, the application of sky flat-fielding will enhance the sensitivity of the VIS detector at scales of larger than $\ang{;;1}$, up to the size of the FOV, enabling \Euclid to detect extended surface brightness structures below $\mulim=31$ \magarc\ and beyond.}

   \keywords{Instrumentation: detectors -- Techniques: image processing -- Space vehicles: instruments -- Techniques: photometric -- Methods: observational -- Galaxies: general}
   \maketitle
%

\section{Introduction}
\label{Sec:Intro}
Deep and wide imaging surveys are the next frontier for many studies in galaxy evolution and cosmology. The study of the structure of stellar halos \citep{Arp1969,Ibata2007,Trujillo2016,Buitrago2017}, the intracluster light \citep{deVaucouleurs1970, Mihos2004, Montes2019, Montes2021}, including the traces of their assembly like tidal tails, shells, and faint satellites, \citep[][]{Zwicky1952, Arp1966, MalinCarter1980, Schweizer1988, Mihos2005, MartinezDelgado2010, MartinezDelgado2015, Bilek2020}, or the detection of the dim ultra-diffuse galaxies \citep{Sandage1984AJ, vanDokkum2018, Trujillo2019} provide critical information about the past evolution of the Universe, and strong tests for the Cold Dark Matter standard cosmological model \citep[$\Lambda$CDM][]{White1978, Bullock2005, Cooper2010, Pillepich2014}. With increasing astronomical image depth, these fields are less affected by the statistical uncertainties of the sky noise and more dominated by systematic biases, such as background gradients, flat-fielding residuals, or the loss of extended sources due to sky over-subtraction, which require special observing techniques and dedicated calibration procedures to recover the full low surface brightness potential of the observatory \citep{Andreon2002,Ferrarese2012, Duc2015, Trujillo2016}. Such effects severely harm the capability of space and ground-surveys to discover and study the structures that are hidden at the very low surface brightness (LSB) limits of the astronomical images. 

\begin{figure*}[th]
 \begin{center}
\includegraphics[width=\textwidth]{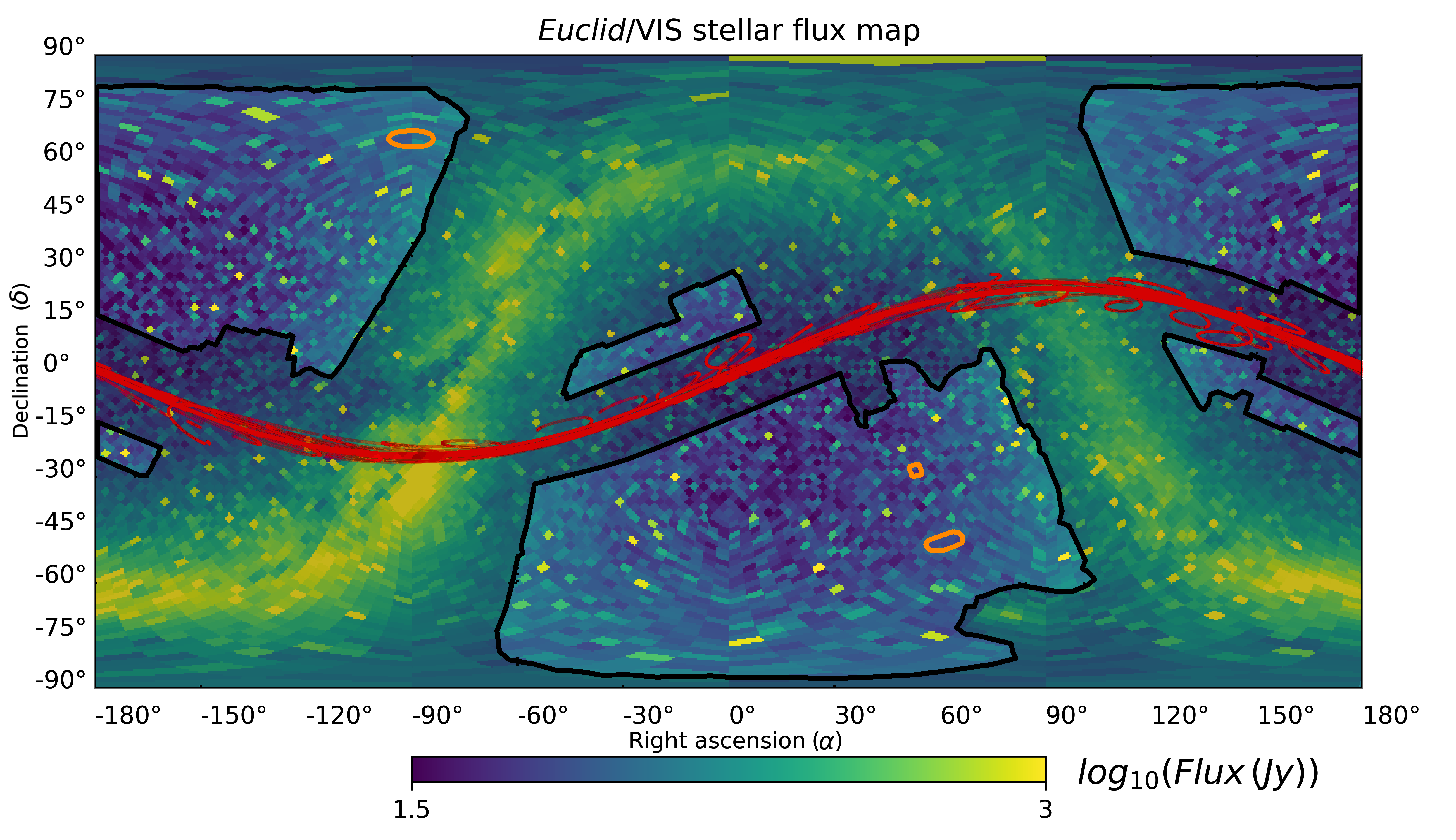}
\caption{\emph{Euclid}/VIS stellar flux density map, based on \emph{Gaia} \citep{GaiaDR1,GaiaDR2} and \citet{Sahlmann2016} catalogs: \emph{Black contours:} Footprint of a proxy of the \emph{Euclid}/VIS Survey \citep{Scaramella2021}, the darkened regions, corresponding to the peak of the Milky Way emission lie outside the footprint. The three regions marked with orange contours correspond to the Deep North, Deep South and Deep Fornax fields. \emph{Color background:} Stellar flux density per HEALpix cell. The brightest region corresponds to the Galactic plane, a region avoided by the \emph{Euclid} footprint. See the bottom colorbar for scaling. \emph{Red lines:} Projected trajectories of the main Solar System bodies (Mercury, Venus, Earth, Mars, Jupiter, Saturn, Uranus, and Neptune) as seen by \emph{Euclid} during the mission, following the ecliptic plane \citep{Giorgini2001}.}
\label{fig:straylight_pointings}
\end{center}
\end{figure*}

Even in space-based observations, one of the most dominant systematic effects in deep cosmological surveys is light gradient contamination \citep[we refer to][for a review on the current challenges in deep imaging]{Mihos2019}. The sky background is a combination of many natural and instrumental effects (i.e, zodiacal light, Earth atmosphere emission, infrared thermal emission, point-spread function contamination and flat-fielding residuals). Space observations present a much lower sky background than ground-based observations, thus increasing the detection capabilities even with lower exposure times. The most common method for background correction is the subtraction of a two-dimensional sky background model with a certain typical variation scale from the image itself  \citep[i.e., Source Extractor,][]{Bertin1996}. While such approach is adequate for the compact source science, these methods are highly sensitive to the accuracy of the fit and the size of the mesh, tending to over-subtract the outskirts of the extended objects, creating regions with artificial negative fluxes around them \citep[we refer the reader to][for a discussion of this effect on the first versions of the Hyper Suprime-Cam Subaru Strategic Program Survey and possible solutions]{Aihara2018}. This effect is particularly common in the mosaics of deep cosmological surveys \citep[see \HST ACS GOODS-North, GOODS-South and WFC3/IR XDF mosaics, ][]{Giacconi2002,Giavalisco2004,Beckwith2006,Koekemoer2013,Illingworth2013} and can also severely affect the detection faint compact sources. Such artifacts result from the sky model inclusion of emission of extended sources like galaxies or cirrus deeply buried in the background noise of the individual images at intensity levels significantly below the $1\sigma$ limit \citep{gnuastro}. If not appropriately masked, the extended source emission can be confused with the background and subtracted. Thus, careful masking of sources and robust statistics are required to avoid over-estimating the sky-background.


\Euclid \citep[][]{Laureijs2011} is a space mission designed to investigate the nature of Dark Energy and Dark Matter through two specific cosmological probes: weak lensing and galaxy clustering, using the  \Euclid\ Visual instrument \citep[VIS,][]{Cropper2014} for optical imaging, and the Near-Infrared Spectrometer and Photometer instrument \citep[NISP,][]{Maciaszek2014}. \Euclid's combination of large survey area (Wide Survey: $15\,000$ deg$^2$, Deep Survey: $40$ deg$^2$, see Fig.\,\ref{fig:straylight_pointings}), high spatial resolution (${\rm FWHM}_{\rm VIS}=\ang{;;0.2}$, ${\rm FWHM}_{\rm NISP}=\ang{;;0.3}$), and depth of both VIS (optical, broad single bandpass $560$--$900$ nm) and NISP (near-infrared - NIR - $Y$, $J$ and $H$) is also ideal for the study of the low surface brightness limits of extended structures, such as Galactic dust cirri, extra-galactic shells and tidal tails, ultra-diffuse galaxies and even the cosmic infrared background (CIB). High spatial resolution reduces the effect of confusion by avoiding source blending, improving the sky background correction and allowing different tracers for low surface brightness structures, like the identification of globular clusters \citep{Montes2020}. In the present article we focus on the VIS detector, whose combination of high-resolution, broadband sensitivity, wide field-of-view (FOV), purely reflective design, and exceptional point-spread function (PSF) stability is highly advantageous for the study of the structure of galaxies. 

The VIS instrument uses calibration lamps to create high-SN flat fields for the correction of the pixel response non-uniformity (PRNU). The lamps illuminate the focal plane directly. The flats are acquired \textit{on sky} as follows: the exposure starts, shutter opening movement lasting three seconds, the lamp illuminates the focal plane (up to two seconds), shutter closes, and the exposure stops. These flats therefore combine the sky background, astrophysical sources, shutter illumination non-uniformity, and the direct illumination by the flat lamp. By dithering the telescope, the astrophysical sources can be removed statistically, and the PRNU can be corrected for with high-precision on spatial scales smaller than 100 pixels. On larger scales, the shutter illumination non-uniformity, the Lambertian cosine law for the calibration lamp, and any intrinsic illumination properties of the telescope optics will results in non-uniform illumination. Hence, the relative photometric zeropoint will vary across the field of view after application of the lamp flat field. These large-scale deviations will be calibrated to within 0.6\%, using widely dithered observations of a stellar field, measuring how the fluxes of stable photometric sources change as a function of position in the focal plane after the lamp flats were applied (``self-calibration'').

While this approach meets the requirements for \Euclid's core science objectives, it can probably be improved upon for legacy science of the low surface brightness Universe, as we investigate in this paper. Sky flat-fields computed from hundreds of individual images \citep{Pirzkal2011} are a challenging but very accurate technique for reducing artificial large-scale background structures following flat correction \citep{Bouwens2011, Brooks2016, ISR_Mack2017}, also in ultra-deep ground-based observations \citep[\mulim$=31.5$ \magarc\ at 3$\sigma$ in 10$\times$10 arcsec$^2$,][]{Trujillo2016}. For instance, the low surface brightness structures around the galaxies of the WFC3/IR \emph{Hubble Ultra Deep Field} (HUDF, see XDF, \citealt{Illingworth2013}; HUDF12, \citealt{Koekemoer2013}) were considerably suppresed by the original reduction process. In \citet{Borlaff2019}\footnote{The ABYSS HST Ultra Deep Imaging Project: http://www.iac.es/proyecto/abyss/}, the authors reduced the systematic biases associated with the reduction process using careful sky flat-fielding and optimized background correction techniques. These methods recovered a great number of new structures on the outskirts of the largest galaxies on the HUDF. As a result of the background improvements, some galaxies now present nearly double the size than in the previous images, showing extended discs and stellar halos while increasing the depth of the images.


Despite that standard imaging pipelines are accurate enough to recover the properties of relatively compact sources, this is not the case for extended low surface brightness imaging. In absence of additional processing, the resulting data compromises the morphology and photometry of any structure with relatively extended spatial scales in the final mosaics. \Euclid's sky-mapping strategy is optimally suited for sky flat fielding. Results can be compared with the internal calibration lamp flat fields and large-scale selfcal measurements, and readily applied to the data as an additional correction if necessary, monitoring possible contamination sources and other unwanted effects on the detectors in real time.  



The questions that arise are, given the characteristics of the \Euclid mission:
\begin{enumerate}
    \item What is the expected surface brightness limiting magnitude for extended sources?
    
    \item Which will be the main contributors to the sky background affecting the low surface brightness performance? 
    
    \item Can we apply specific reduction techniques to obtain high-quality mosaics that preserve the properties of extended low surface brightness sources?
    
    \item How can we efficiently predict the presence and structure of unwanted stray-light contamination?
    
    \item Is sky flat-fielding a valid strategy to calibrate the variation in sensitivity across the FOV?
\end{enumerate}

By analyzing these questions we explore the efficacy of \Euclid for low surface brightness science. In the present work, we generate $9916$ VIS image simulations (enough to study the precision of sky-flat fields over an extended period of time, which approximately correspond to the first 4 months into the mission) with the main objective of assessing the deep imaging capabilities of the survey. The paper is organized as follows: We describe the process to generate the realistic VIS simulations in Sect.\,\ref{Sec:Methods}. Section \ref{Sec:results} is dedicated to the description of the results. Sections \ref{Sec:Discussion} and \,\ref{Sec:Conclusions} contain the discussion and conclusions respectively. All magnitudes are in the AB system \citep{Oke1971} unless otherwise noted. 

\section{Methods}
\label{Sec:Methods}

\Euclid will be located in a Lissajous orbit in the Sun-Earth Lagrangian point L2. In this orbit, the optical/NIR background is mainly a combination of the zodiacal light, stray-light from stars and Solar System bodies, the CIB and the interstellar medium (ISM) of the Milky Way. We must note that our objective is not to eliminate these components, but to be able to identify and separate them. If these components do not create a significant gradient (we test this in Sect.\,\ref{Subsec:results_stray}), we can assume that such sky background is a dim, but naturally flat illuminating source, that theoretically should allow us to calibrate an imaging detector from variations in the pixel-to-pixel sensitivity across the field of view \citep{Chromey1996}. This technique is called sky flat-fielding and provides the sensitivity correction of the detector using the science exposures themselves. 

Our purpose is to evaluate if the sky background seen by VIS at L2 has a SNR sufficient to create a flat-field correction that does not increase the noise of the final mosaics, and how many co-added science exposures are needed to obtain a reasonable calibration. At the same time, we want to test if there will be systematic stray-light gradients and how they could affect this correction. In Sect.\,\ref{Subsec:methods_skybackground} we detail the results from our simulated observations of the zodiacal light, CIB and ISM as seen by the VIS detectors. The stray-light component and the evaluation of its gradients will be addressed in Sect.\,\ref{Subsec:methods_straylight}.

\subsection{Sky background simulation}
\label{Subsec:methods_skybackground}

\begin{figure*}[th!]
 \begin{center}
\begin{minipage}[c]{9.08cm}
\includegraphics[width=\textwidth, trim={0 0 0 0},clip]{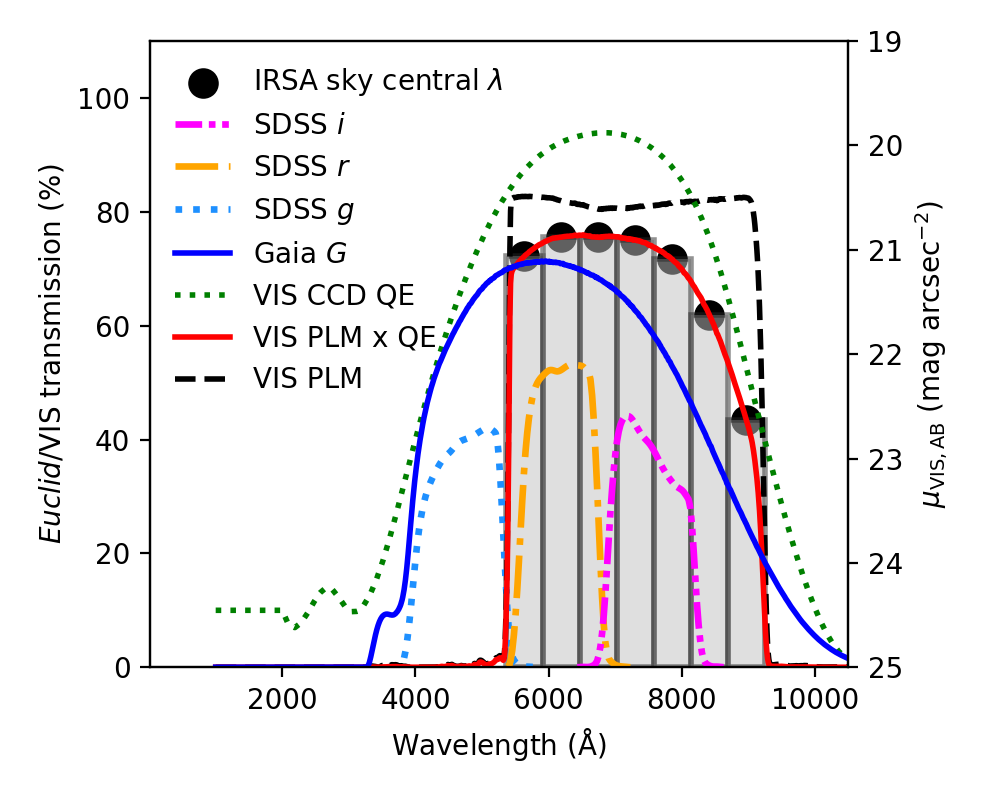}
\end{minipage}
\begin{minipage}[c]{9.23cm}
\includegraphics[width=\textwidth, trim={0 0 0 0},clip]{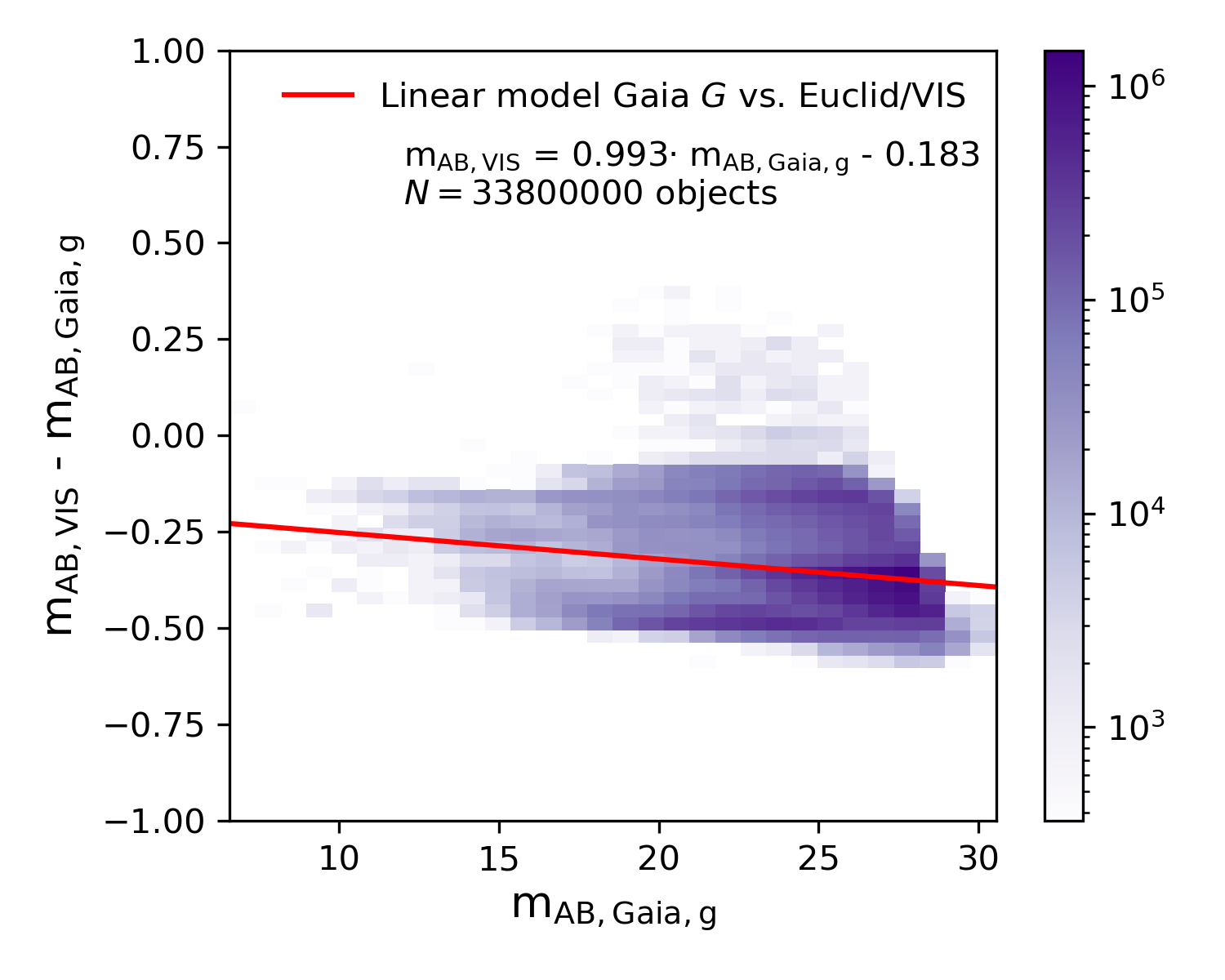}
\end{minipage}
\caption{\emph{Left panel:} \emph{Solid red line:} \emph{Euclid}/VIS transmission curve. \emph{Solid blue line:} Gaia $G$-band transmission curve \citep{Venancio2020}. \emph{Black dashed line:} \emph{Euclid}/VIS payload module (PLM) transmission curve. \emph{Green doted line:} Quantum Efficiency of the \emph{Euclid}/VIS CCD detectors (including End of Life contamination), derived from initial engineering models. \emph{Grey columns:} Wavelength ranges for the numerical integration of the transmission of VIS using the IRSA sky background model. \emph{Light-blue dotted line:} SDSS $g$-band sensitivity for extended sources with zero airmass. \emph{Orange dash-dotted line:} Equivalent for SDSS $r$-band. \emph{Magenta dash-dot-dotted line:} Equivalent for SDSS $i$-band. \emph{Right panel:} \emph{Euclid}/VIS magnitude as a function of the Gaia $G$-band AB magnitudes for the synthetic stellar objects of the \emph{Euclid} True Universe simulation. See the legend for the fitted linear transformation model between both bands.}
\label{fig:IRSA_VIS_spectra}
\end{center}
\end{figure*}

The fraction of the sky background dominated by the zodiacal light, stray-light, ISM, and the CIB in L2 is strongly dependent on the position on the sky and also with time, especially in the case of the zodiacal light. As a consequence of this, the \emph{Euclid} survey avoids bright stars, and those regions of maximum zodiacal light \citep[see][]{Tereno2014, Scaramella2021}. Zodiacal light and stray-light from stars are the dominant components of the sky background in the optical and NIR region of the spectrum. This background level increases the noise of the images, but also provides a useful reference uniform light component to create large-scale sky flats. Thus, creating realistic simulations of the sky background is the key point in the present study. 

In order to develop a realistic sky model, we take advantage of the background model calculator provided by the NASA/IPAC Infrared Science Archive (IRSA)\footnote{NASA/IPAC Infrared Science Archive Background Model https://irsa.ipac.caltech.edu/applications/BackgroundModel/}. The IRSA background model provides estimations based on observations for the different sky background components considered in this study (zodiacal light, stray-light, CIB, and ISM), as a function of the observation time (day of the year), observation wavelength (from $0.5$ to $1000$ \micron), and the sky coordinates. We refer to the project webpage for details on how the different components of the sky background are modeled. A table-based query system allows the user to calculate the spectral brightness (MJy sr$^{-1}$) at the required pointing, wavelength, and time of the year\footnote{We note that this sky-background model is independent from that of other \emph{Euclid} projects and its results may have some differences with those presented in other papers from the collaboration}. 

To estimate the flux that will be detected by the VIS detectors, we numerically integrate the sky background intensity for all the pointings of the \Euclid/VIS Survey footprint from $5640\,\AA$ to $9000\,\AA$, using seven sub-bands of $556\,\AA$ width each (see left panel of Fig.\,\ref{fig:IRSA_VIS_spectra}). The intensity of each bandpass is multiplied with the expected value of instrument response at the central wavelength to obtain the observed spectral energy distribution (SED) of each sky background component. The instrument response ($e$) combines the effects of obstruction, mirrors absorption, dichroic reflectivity, and the quantum efficiency (QE) curve for the VIS CCD detectors, and it is defined as the flux ratio detected by the instrument (VIS) and that received at the entrance of the telescope, as a function of wavelength.

We can calculate the sky background surface brightness (\muvis) in AB magnitudes per arcsec$^2$ as it will be detected with the VIS instrument as
\begin{equation}
\label{eq:VIS_mag_integration_2}
\muvis = -2.5 \log_{10}\Bigg({\frac{\int f(\nu) \, (h\,\nu)^{-1} \, e(\nu) \, {\rm d}\nu}{A \int  (h\,\nu)^{-1} \, e(\nu)\,{\rm d}\nu}}\Bigg) + 8.90 \,,
\end{equation}

\noindent where $f(\nu)$ is the flux measured at a certain central frequency $\nu$, $e(\nu)$ the corresponding instrument response at the same frequency, and $A$ the angular area of the VIS pixels. Following a numerical integration over spectral bandpass bins ($i$), and taking into account the sky background intensity at the center of the selected bandpass ($\nu_i$) and the spectral widths ($\Delta\nu_i$), we can assume a discrete integration over the bins ($i$) defined on the VIS transmission curve
\begin{equation}
\label{eq:VIS_mag_integration_2}
\muvis = -2.5 \log_{10}\Bigg({\frac{\sum f(\nu_{i}) \,  (h\,\nu_{i})^{-1} \, e(\nu_{i}) \, \Delta\nu_i}{A \sum  (h\,\nu_{i})^{-1} \, e(\nu_{i})\,\Delta\nu_i}}\Bigg) + 8.90 \,.
\end{equation}
The results from this analysis are detailed in Sect.\,\ref{Subsec:results_skybg}.
\subsection{Stray-light contamination}
\label{Subsec:methods_straylight}

In this section we describe the process to simulate the stray-light from stars, including both stars in and out of the FOV. For \Euclid/VIS, the stray-light is expected to be the second most important contributor to the sky background level after zodiacal light. The broad term of stray-light comprises any flux that does not belong to the on-axis source, which is usually the object of interest, either point-source or extended. If thermal isolation, baffles and the rest of the optical components of the telescope were ideal, creating no significant scattering or secondary optical paths, and if there were no diffraction effects, the photons collected at a single pixel would only be originated at the source located in the on-axis line-of-sight from that point in the FOV. 

To simulate real-world observations, we can divide stray-light sources in two different types \citep{DesignConstructionLargeTelescopes, Lemke2003, Spangelo2015}:
\begin{enumerate}
    \item Sources outside the line-of-sight, either astronomical or not, such as stars, planets, the Moon or the Earth,
    
    \item Thermal emission from the telescope components that surround the detectors.
\end{enumerate} 
In the case of \emph{Euclid}, the background generated by the VIS equipment's thermal emission is estimated to be $1.52\times10^{-28}$ e$^-$ px$^{-1}$ s$^{-1}$ (this estimation is based on an internal ESA study with the support of industry), thus deemed negligible for this study. 

Specular and scattered light from off-axis light in the optical components contribute to the background level in the images, increasing the noise. Stray-light contamination is one of the most important factors to take into account in the observation planning phase. This work presents a similar stray-light analysis as  \citet{KlaasOkumura2014} for the PACS and SPIRE instruments of \emph{Herschel} telescope. The function that defines external stray-light transmission of a telescope is the Normalized Detector Irradiance \citep[NDI hereafter,][]{DesignConstructionLargeTelescopes}. The NDI is defined as the ratio between the stray-light irradiance (power per unit area) at the detector to the irradiance of the source at the entrance of the telescope, allowing to estimate the flux of photo-electrons that an off-axis source will generate on a certain region of the detector. This function is strongly dependent on the optical setup and wavelength. For a given telescope, the NDI depends on the angular distance between the optical axis and the source ($\theta$), the position angle of the source in the focal plane reference frame ($\phi$), the observation wavelength ($\lambda$) and the position on the FOV ($x,y$),

\begin{equation}
\label{eq:NDI_definition}
{\rm NDI}(\theta,\phi,\lambda,x,y) = \frac{E_{\rm stray}(I, \theta, \phi, \lambda, x, y)}{E_{\rm source} (I, \lambda)} \,.
\end{equation}

As a consequence of the complex dependence of the NDI on the specific characteristics of the detector, the optical system and the sources, its solution is usually numerically estimated through ray-tracing simulations and realistic 3D models of the system. In the case of the \emph{Euclid}/VIS detector, two models are available for the NDI \citep{GasparVenancio2016}. First, an NDI model was created using the stray-light analysis software ASAP \citep{Turner2004}. This model considers the variation of the NDI with the distance to the source and orientation of the detector ($\theta$, $\phi$), and its dependence within nine different positions across the focal plane of VIS (F1--F9, see Table \ref{table:NDI_envelope_parameters}). It is important to note that the current NDI models do not include diffraction peaks or ghosts created by the secondary reflections on the optical elements. Updated estimations of the NDI which include this component will be presented in a forthcoming publication. The non-axisymmetric NDI model has been calculated for a finite number of $\theta$ and $\phi$ positions, thus numerical interpolation is required to estimate the NDI at each position and subsequently, the stray-light contamination from them. 

A second and simplified version of the VIS NDI model (worst case scenario or envelope NDI model) was created choosing the higher NDI level of all the position angles ($\phi$) at a certain angular distance from the optical axis to the source ($\theta$). This model depends only on $\theta$ (hence one-dimensional), therefore it does not accurately represent the directional baffling effect of the telescope optics. This simplified model can be approximated using the following set of equations (we refer to Table\,\ref{table:NDI_envelope_parameters} for the definition of the different parameters of this expression and their dependence across the FOV):

\begin{equation}
\label{eq:NDI_envelope}
{\rm NDI}(\theta, \lambda) = A(\theta, \lambda) \frac{1}{1 + \Big(\theta / \theta_{\rm 1s}\Big)^2} \frac{1 + \Big(\theta / \theta_{\rm 2e} \Big)^2}{1 + \Big(\theta / \theta_{\rm 2s}\Big)^4} \,,
\end{equation}
where:
\begin{equation}
\label{eq:A_NDI_envelope}
A(\theta, \lambda) = A(0, 550)\,\Bigg(\frac{\lambda}{550\,\mathrm{nm}}\Bigg)^{n(\theta)}\,,
\end{equation}
with $\lambda$ being the stray-light source wavelength in nm and
\begin{equation}
\label{eq:n_NDI_envelope}
n(\theta) = -1.8\,\frac{1}{1 + \Big(\theta / \theta_{\rm wd1}\Big)^{0.75}} \frac{1}{1 + \Big(\theta / \theta_{\rm wd2}\Big)^{20}} \,.
\end{equation}

\begin{table}
{\small 
\begin{center}
\caption{Parameters of the envelope NDI model defined in Eqs.\,(\ref{eq:NDI_envelope}-\ref{eq:n_NDI_envelope}) \citep{GasparVenancio2016}. Rows 1--3 contain the values for $\theta_{\rm 1s}$, for the corners and center of the FOV (F1--F9). Same for A(0,550) in rows 4--6. Row 7 contains the constant parameters $\theta_{\rm 2s}$, $\theta_{\rm 2e}$, $\theta_{\rm wd1}$ and $\theta_{\rm wd2}$.}
\begin{tabular}{ccccc}
\toprule
\textbf{$\theta_{\rm 1s}$} & \diagbox{Y}{X} & $\textbf{\ang{-0.390;;}}$ & $\textbf{\ang{0.0;;}}$ & $\textbf{\ang{0.392;;}}$ \vspace{0.1cm}\\
1) & $\textbf{\ang{0.47;;}}$ & 0.020 & 0.025 & 0.031 \\
2) & $\textbf{\ang{0.82;;}}$ & 0.017 & 0.020 & 0.024 \\
3) & $\textbf{\ang{1.17;;}}$ & 0.013 & 0.015 & 0.017 \\
\midrule
A(0,550) &  \diagbox{Y}{X}  & $\textbf{\ang{-0.390;;}}$ & $\textbf{\ang{0.0;;}}$ & $\textbf{\ang{0.392;;}}$\vspace{0.1cm}\\
4) & $\textbf{\ang{0.47;;}}$ & 126 & 84 & 54 \\
5) & $\textbf{\ang{0.82;;}}$ & 177 & 126 & 90 \\
6) & $\textbf{\ang{1.17;;}}$ & 300 & 240 & 180 \\
\midrule
7) & $\theta_{\rm 2s}=\ang{15;;}$& $\theta_{\rm 2e}=\ang{35;;}$& $\theta_{\rm wd1}=\ang{0.3;;}$ & $\theta_{\rm wd2}=\ang{2;;}$\\
\bottomrule
\vspace{0.2cm}
\end{tabular}

\label{table:NDI_envelope_parameters}
\end{center}
}
\end{table}

Once we have the NDI for a certain source, as a function of $\theta$, $\phi$, and its position on the FOV, we can simulate the stray-light contamination ($S$, e$^-$\,\si{{\rm px}^{-1}\,{\rm s}^{-1}}) created by a source of magnitude $m_{\rm AB}$, that produces an irradiance ($I$, \si{W m^{-2}}) at the entrance of the telescope as
\begin{equation}
\label{eq:straylight_estimation_3}
S(I,\theta,\phi,\lambda,x,y) = {\rm NDI}(\theta,\phi,\lambda,x,y)\,I\,A\,T\, \frac{\lambda_{\rm ref}}{h\,c} \,,
\end{equation}
where $h$ is the Planck constant (\si{kg m^2 s^{-1}}), $c$ is the speed of light (in \si{m s^{-1}}), the reference bandpass wavelength is $\lambda_{\rm ref}=7.25 \times 10^{-7}$ m, $T$ is the average VIS transmission (which is approximately 76\% at the peak of the curve), $A$ is the physical pixel area expressed in m$^2$ ($1.44\times10^{-10}$ m$^2$ for \Euclid/VIS) and
\begin{equation}
\begin{aligned}
\label{eq:straylight_estimation_1}
I = \int_{\lambda_{\rm min}}^{\lambda_{\rm max}} f_{\nu} \frac{c\, {\rm d}\lambda}{\lambda^2} \approx c f_{\nu} \frac{\lambda_{\rm max} - \lambda_{\rm min}}{\lambda_{\rm max}  \,\lambda_{\rm min}} \,.
\end{aligned}
\end{equation}

\noindent From the AB magnitude equation in units of \si{W m^{-2} Hz^{-1}}, we can define the spectral flux density ($f_{\nu}$) as
\begin{equation}
\label{eq:straylight_estimation_2}
f_{\nu} = 10^{-0.4(m_{\mathrm{AB}}+56.1)} \,. 
\end{equation}


\begin{figure*}[t]
 \begin{center}
\includegraphics[width=\textwidth, trim={40 30 40 60},clip]{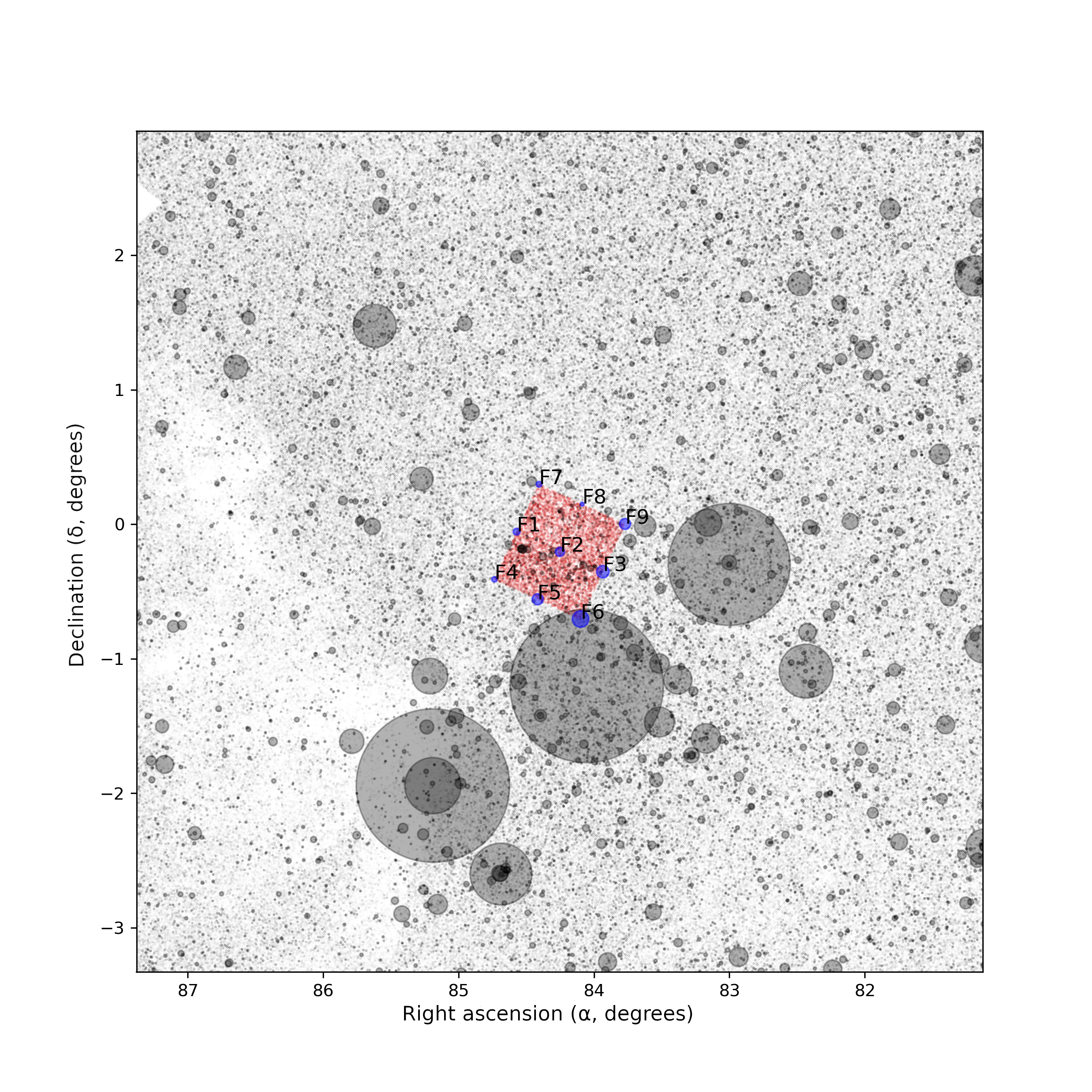}
\caption{Stray-light simulation of a \emph{Euclid}/VIS observation in the environment of Orion's belt ($\alpha = \ang{84.054;;}, \delta = \ang{-0.202;;}$). \emph{Grey transparent circles:} Stars outside the VIS focal plane and at $R<\ang{5;;}$ from the center of the FOV. The radius of each circle is log-scaled to the flux of the star in the VIS band. The three largest circles represent the main stars  $\zeta$ Ori, $\eta$ Ori, and $\delta$ Ori, from left to right, respectively. \emph{Red points:} Stars inside the FOV of VIS. \emph{Purple circles:} Stray-light contamination level (infield and outfield) at the F1--F9 focal plane points. Note how the focal plane points closer to the bright stars are gradually more contaminated. To the East ($\alpha \sim \ang{87;;})$, dust extinction from NGC2024 and Barnard 33 nebulae are visible, diminishing the brightness in the stars in the background. Note that the field was chosen for illustrative purposes, as \Euclid\ surveys will not observe these regions deep into the Galactic plane.}
\label{fig:NDI_num_example}
\end{center}
\end{figure*}

In order to simulate the stray-light produced by stars in the \emph{Euclid} Survey, we use the \emph{Gaia} Catalog \citep{GaiaDR1,GaiaDR2}. The \emph{Gaia} catalog has 10$^9$ sources, including broadband photometry in the $G$ band with a faint limit of $G = 21$ mag and a bright limit of $G\sim6$. We complement the \emph{Gaia} DR2 Catalog with the additional catalog of 230 bright stars ($G<6$ mag) from \citet[][]{Sahlmann2016}. To transform from \emph{Gaia} $G$-band to \emph{Euclid}/VIS fluxes, we calibrate a transformation model using the synthetic catalogs from the \emph{Euclid} True Universe Simulation\footnote{\Euclid Flagship simulation: https://www.euclid-ec.org/?page\_id=4133} (paper in preparation, see right panel of Fig.\,\ref{fig:IRSA_VIS_spectra}). The \emph{Euclid} True Universe simulation provides synthetic photometry for $4.1\times 10^{7}$ stars in $1.8\times 10^{4}$ deg$
^2$ combining the stellar population models of \citet{Pickles2010} for the bright end and of the Besançon galaxy model\footnote{Besançon Model of the Galaxy web site: https://model.obs-besancon.fr/} \citep{besancon} for the faint end of the luminosity distribution. Finally, we include the stray-light emission from the major Solar System bodies, taking into account their predicted sky position as a function of time as seen from L2 by \emph{Euclid}, based on the NASA/JPL HORIZONS ephemeris\footnote{NASA/JPL HORIZONS Online Ephemeris System: https://ssd.jpl.nasa.gov/?horizons} \citep{Giorgini2001}. 

Integrating the stray-light created by $\sim10^9$ independent sources in several different positions of the FOV is a challenging computational task. In order to optimize the process, we have adopted an approximation for the objects beyond a certain angular distance from the center of the FOV. We define a certain high-resolution region surrounding the center of the \emph{Euclid}/VIS FOV ($R<5\degree$) where we calculate the stray-light from every star individually. Outside that region, the sky is divided in a grid of HEALPix\footnote{HEALPix is a project of NASA Jet Propulsion Laboratory available at: https://healpix.jpl.nasa.gov/} cells \citep{Gorski2005} of approximately the same area. We adopt a characteristic parameter of $N_{\rm side}=32$, which is equivalent to dividing the sky sphere in 12\,288 HEALpix cells and an approximate spatial resolution of \ang{1.8;;}. We show the complete stellar flux map along with the \emph{Euclid}/VIS footprint in Fig.\,\ref{fig:straylight_pointings}. Every star which belongs to a cell located at $R>5\degree$ from the center of the FOV is grouped with the rest of the stars inside the same cell and their flux is estimated as a single source. The position of the group is calculated as the flux weighted mean of the individual positions of the stars. In Appendix \ref{Appendix:limitations_straylight} we provide a quantitative test to the precision of this method, where we define the optimization of the high-resolution limit at $R_{\rm min}=5 \degree$. We find the stray-light estimation converges exponentially with $R_{\rm min}$, obtaining a variation at $\ang{5;;}$ of 0.1--0.01 e$^-$ per exposure per each degree that we increase $R_{\rm min}$. We can conclude that assuming $R_{\rm min}=5 \degree$ provides a high confidence level to the stray-light estimation at an acceptable cost of computational effort. 

We provide an example of our simulations in Fig.\,\ref{fig:NDI_num_example}. We simulate the first $9916$ pointings of the mission plan, taking into account their sky position angle. For each simulation, we find all the HEALpix cells closer than $R_{\rm min}=5 \degree$ to the center of the FOV. Then we generate a new catalog, combining the individual positions and fluxes of the closest stars (high-resolution map) with the positions and fluxes of the HEALpix cells for the sources at $R > R_{\rm min}$. Finally, based on the relative distance, position angle and magnitude of each object in these new hybrid catalogs, we estimate the total stray-light that each star produces at the F1--F9 characteristic focal plane points (see Table \ref{table:NDI_envelope_parameters}), following Eqs.\,(\ref{eq:NDI_envelope} -- \ref{eq:straylight_estimation_3}), and the numerical estimations from the non-axisymmetric NDI model. The results of the stray-light analysis, for both NDI models are presented in Sect.\,\ref{Subsec:results_stray}.

\subsection{Extended source masking}
\label{Subsec:methods_masking}

In this section we detail the methods used to describe the masking of extended sources on our images, one of the most important points in order to accurately simulate the sky flat-fields. All science exposures used to create the sky flats will present astronomical sources. To avoid inhomogeneities in the final sky flats, all objects that are not part of the uniform sky background should be masked. This process decreases the number of valid pixels for the analysis, systematically reducing the precision in detector sensitivity prediction.

\begin{figure}[t]
 \begin{center}
\includegraphics[width=0.5\textwidth, trim={20 20 20 20}, clip]{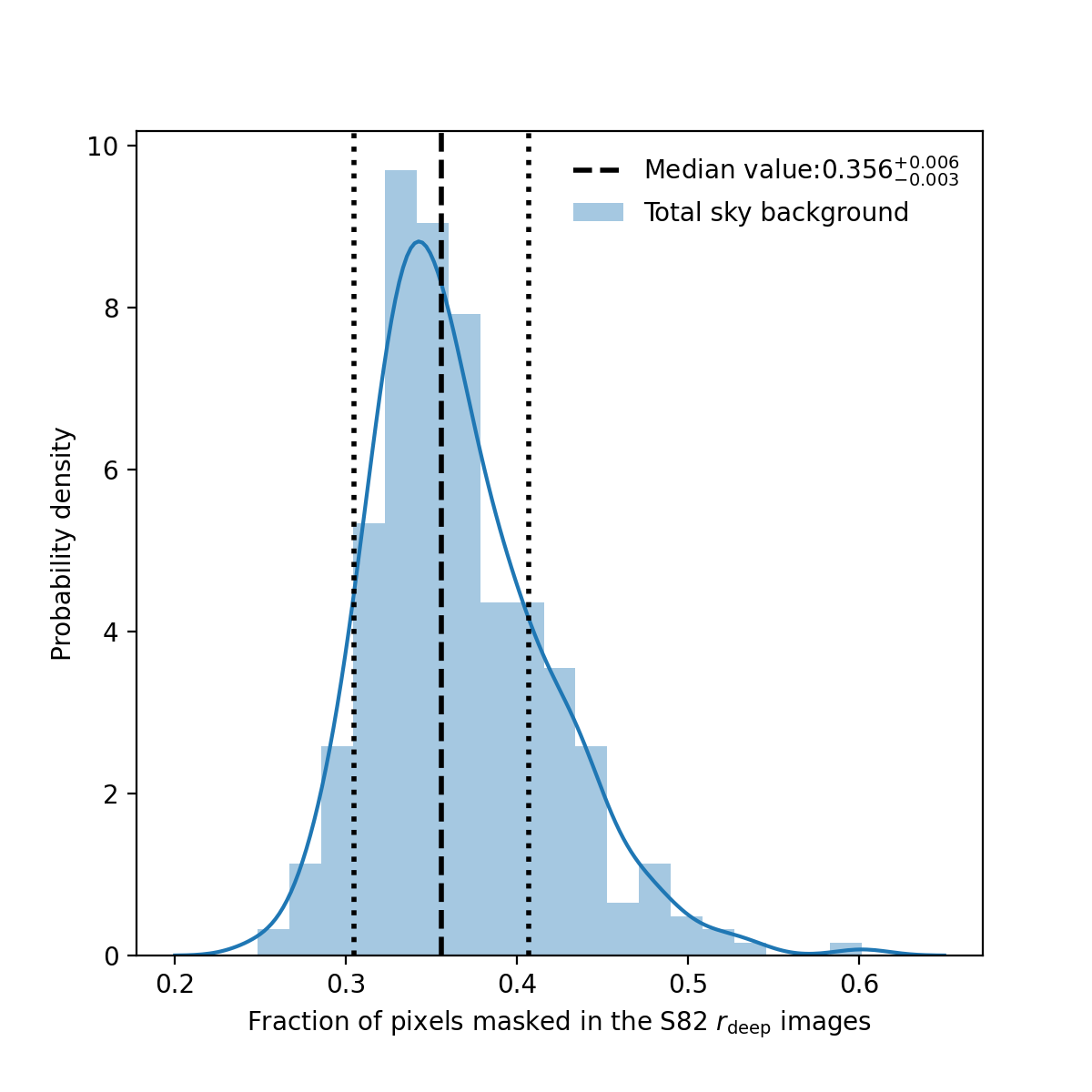}
\caption{Fraction of pixels identified and masked as part of a source, per exposure in the IAC SDSS S82 mosaics \citep{IACStripe82,Roman2018}. The dashed and dotted vertical lines represent the median value of the distribution and its $\pm1\sigma$ dispersion percentiles (see the legend).}
\label{fig:filling_fraction}
\end{center}
\end{figure}

To create a realistic distribution of sources obtained with a similar depth in a region of the spectrum, we make use of one of the most extended and deep surveys available for low surface brightness, the IAC SDSS Stripe 82 (S82) Legacy Survey \citep{IACStripe82,Roman2018}. The S82 is a 275 deg$^2$ region along the celestial equator ($-50\degree<\alpha<+60\degree, \ang{-1.25;;}<\delta<\ang{+1.25}$) which has been repeatedly observed during the SDSS Survey \citep{York2000}. Each region of the S82 has been observed approximately 80 times, providing a limiting surface brightness $2.4$ \magarc\ fainter than that of standard SDSS data. The authors carefully corrected for residual sky background substructures that might bias low surface brightness structures. The mosaics were generated using $u, g, r, i,$ and $z$ filters, plus an additional mosaic denominated $r_{\rm deep}$, which combines the deepest frames of the $g, r,$ and $i$ bands into single mosaics. These frames are dedicated for the detection of extended low surface brightness structures, particularly suitable for our work, since the maximum VIS sensitivity range overlaps well with the combined $r_{\rm deep}$ SDSS synthetic band (see Fig.\,\ref{fig:IRSA_VIS_spectra}). In addition, the bands selected for the $r_{\rm deep}$ mosaics present the deepest limiting surface brightness magnitudes ($\mu_{\rm lim,S82}=29.1, 28.6, 28.1$ \magarc). As we will detail in Sect.\,\ref{Subsec:results_limitingmagnitude}, the depth of these observations is compatible with the expected surface brightness limiting magnitude in the VIS exposures ($\mu_{\rm lim,S82}=29.5$ \magarc\ per field in the Wide Survey, measured at a $3\sigma$ level, over $10\times10$ arcsec$^2$).

\begin{figure*}[t!]
 \begin{center}
\includegraphics[width=0.48\textwidth]{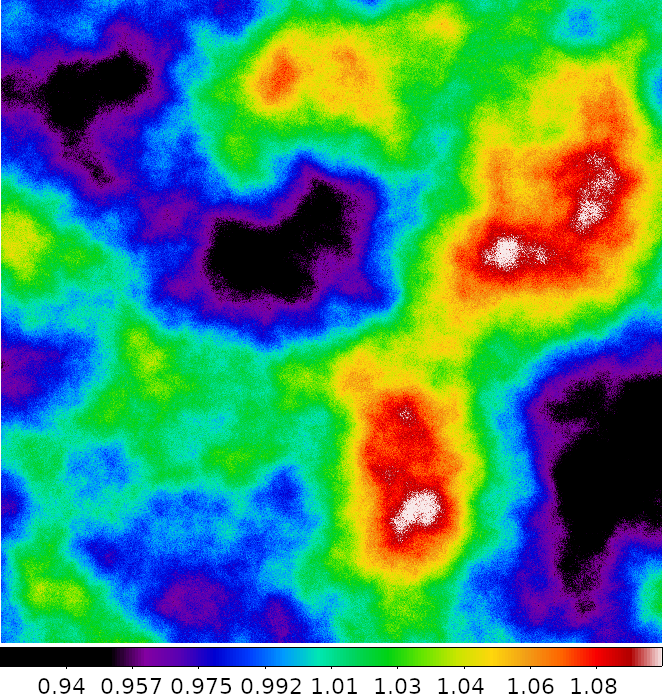}
\includegraphics[width=0.48\textwidth]{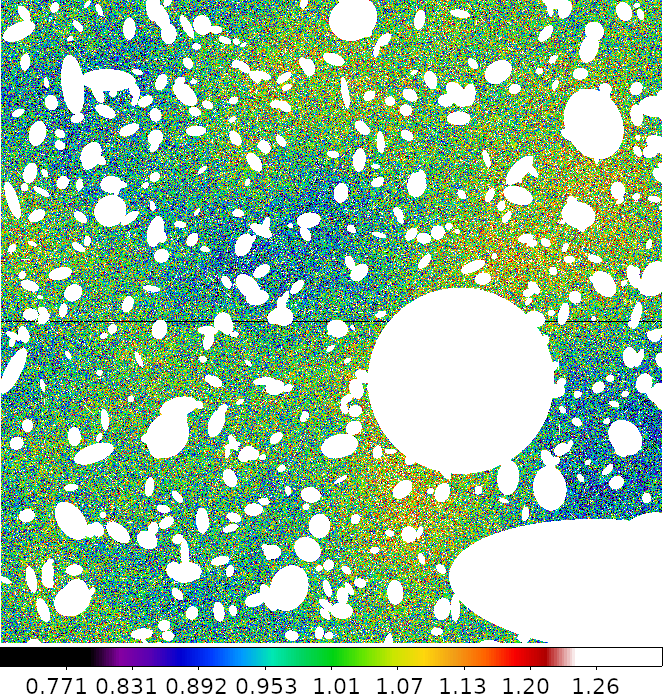}
\caption{\emph{Left panel:} Simulated flat-field structure using Perlin and pixel-to-pixel Gaussian noise. \emph{Right panel:} Example of one of the simulated pre-calibrated frames ($\ang{;6.83;}\times\ang{;6.89;}$) used for the construction of the sky flats. White regions represent the source masks. Bias and Dark corrections have been applied, but no flat-field correction has been performed, showing the same background structure than the original flat-field in the left panel. See the colorbar in the bottom.}
\vspace{-0.2cm}
\label{fig:skysim_example}
\end{center}
\end{figure*} 

Due to the low Galactic latitude of some regions of the S82 Survey, the number of stars and Galactic cirrus is notably larger than in typical \emph{Euclid}/VIS exposures. In fact, the S82 is located at the edge of the \emph{Euclid} Wide Survey footprint. The increased source count will systematically bias our results to a larger number of masked objects and consequently lower statistics for the sky flat-fields. In addition, the lower resolution of the S82 images compared to that of \emph{Euclid} increases the size of the masked regions. For our purposes, we assume that these conditions represent the worst-case scenario for this calibration, and that we will obtain better statistics in the real \emph{Euclid} sky flats. Despite these considerations, the wavelength coverage and depth of the IAC S82 Legacy Survey are optimal to simulate the number of pixels that will be covered by sources in average exposures. 

In order to study the apparent size and basic morphology of the objects in the $r_{\rm deep}$ images, we use {\textsf{Gnuastro}}'s {\tt{Noisechisel}} and {\tt{Segment}} utilities \citep{gnuastro, noisechisel_segment_2019}. To optimize the detection of the faintest wings of the extended sources, we set {\tt{tilesize}} to $70\times 70$ pixels$^2$ and the minimum number of neighbors for interpolation to 3 in {\tt{Noisechisel}}. For a more detailed description we refer to the {\textsf{Gnuastro}} tutorial to detect large extended targets\footnote{{\textsf{Gnuastro}} Tutorial - Detecting large extended targets: https://www.gnu.org/s/gnuastro/manual/html\_node/\\Detecting-large-extended-targets.html}. Using the source-detection maps, we measure the fraction of pixels that belong to a detectable source in each image (the filling factor). Approximately 30--40\% of the pixels were identified as part of a source (see Fig.\,\ref{fig:filling_fraction}). Using the source-detection maps, we generate a catalog recording the area, major axis size, ellipticity and position angle of all detected sources. We transform the major axis sizes from the SDSS pixel scale to the \emph{Euclid}/VIS pixel scale.

Once the source catalogs are generated, the process to create the masks for each simulated pointing can be summarized as follows: 
\begin{enumerate}
 
    \item We select a filling factor following the observational probability distribution (see Fig.\,\ref{fig:filling_fraction}). 
    
    \item We select random sources until the sum of their equivalent areas on the CCD is equal to the required number of pixels to be masked, set by the chosen filling factor. 
    
    \item The masks are placed randomly such that a certain fraction of them overlap. The overlapping areas systematically reduce the real amount of pixels masked in each  simulation. To partially compensate for the reduction of masked pixels, we generate a randomly placed single circular mask equal to the net area of mask-overlap. 
    
    \item Even after this correction, some of the objects will overlap with the compensating circular mask. As a last step, we mask additional random pixels until we reach the required filling factor for the simulation. 
\end{enumerate}

Finally, we simulate the effect of cosmic rays (CR) in the images by using the CR generation module of the {\tt{VIS-PP}}\footnote{The \emph{Euclid} Visible InStrument Python Package (VIS-PP) was created by Sami-Matias Niemi and it is available through {\tt{GitHub}}: https://github.com/sniemi/EuclidVisibleInstrument} Python package for \emph{Euclid}/VIS simulations. CRs are added until they cover 2\% of the FOV, a worst case value considered in the technical requirements. We present an example of the resulting masks in the right panel of Fig.\,\ref{fig:skysim_example}, with a completed sky background simulation (see Sect.\,\ref{Subsec:methods_VIS_exposure_simulator}).

\subsection{Flat-field}
\label{Subsec:methods_flatfield}

In order to estimate our precision to recover the structure of the VIS detector sensitivity using the sky background, we need to include the effects of a realistic photo response non uniformity (PRNU) in our simulations. Furthermore, the effective system transmission can be modified by molecular contamination, mostly in form of water ice on optical surfaces due to molecular outgassing. This is a common problem encountered by spacecraft and can easily change photometry by up to several ten percent \citep[i.e., \emph{Gaia}, ][]{Gilmore2018}. Most contamination can be cleared by heating of the optical surfaces when necessary. Unlike the Euclid lamp flats, sky flat-fields take into account the full optical path, which can produce a significantly different calibration in the case of surface contamination. Therefore, we must take into account a certain time variation of this sensitivity.

According to the \emph{Euclid} payload element requirements, the VIS instrument pixel-to-pixel relative response is predicted to be stable to better than $10^{-4}$ on a 24 hours time-scale, and better than $2.5\times10^{-3}$ on a monthly time scale, when averaged over $100\times100$ pixels$^2$. Assuming the worst-case scenario based on these requirements, we can generate a function that simulates a realistic sensitivity for the VIS CCD including their expected variation with time. In order to do develop this sensitivity function, we follow these steps:

\begin{enumerate}
    \item First, we generate an initial flat-field, which will be the sensitivity at the start of the mission. To generate a realistic complex pattern with variations at different spatial scales, we make use of self-similar (fractal) noise function of the {\textsf{Perlin-numpy}} package\footnote{https://github.com/pvigier/perlin-numpy}. {\textsf{Perlin-numpy}} is an implementation of the {\textsf{simplex}} Perlin noise algorithm presented in \citet{perlin1985image} and later improved in \citet{Perlin2002}. Using a combination of several layers of noise, this algorithm simulates the effect of fractal noise. We normalize the resulting structure to have an average value equal to one, with a minimum-to-maximum amplitude in all the FOV of 0.2 (20\%). Note that this amplitude is arbitrary and does not affect our final results.  
    
    \item To ensure the pixel-to-pixel complexity, we add a pattern of random Gaussian noise with $\sigma=10^{-2}$ (the expected pixel-to-pixel variation). The result is the simulated PRNU at the initial mission time (the ``base flat'', see the left panel of Fig.\,\ref{fig:skysim_example}). 
    
    \item To simulate the time variation of the flat-field without increasing the pixel-to-pixel standard variation, we multiply the base flat with two frames to take into account the small and large-scale time variation. The first is a random Gaussian noise field on a per pixel basis, with a standard deviation $\sigma=10^{-4}\,t$, with $t$ being the mission time in days. Secondly, we include the large-scale variation with a different Perlin noise pattern, spatially smoothed with a Gaussian kernel of $100$ pixels in size, and an amplitude equal to $\sigma=8.2\times 10^{-5}\,t$. We will refer to these components as delta PRNUs.  
\end{enumerate}

Based on the \Euclid VIS payload requirements, very small variations are expected in periods of several days, being almost negligible from exposure to exposure. To account for variations over these timescales, and avoid to artificially increase the noise linearly with time, we generate five delta PRNUs, simulating changes in flat expected in periods of 30 days. We independently multiply each one of these delta PRNUs to the base flat, obtaining five different flats (one every 30 days, for a period of four months). Finally, to estimate the flat-field at a certain mission time, we perform a linear interpolation between the two closest estimations in time. By doing so, the PRNUs will present compatible noise levels, but it will have a difference in structure. 



\subsection{VIS exposure simulation process}
\label{Subsec:methods_VIS_exposure_simulator}

\begin{figure*}[t!]
 \begin{center}
\includegraphics[trim={0 0 0 0}, clip, width=\textwidth]{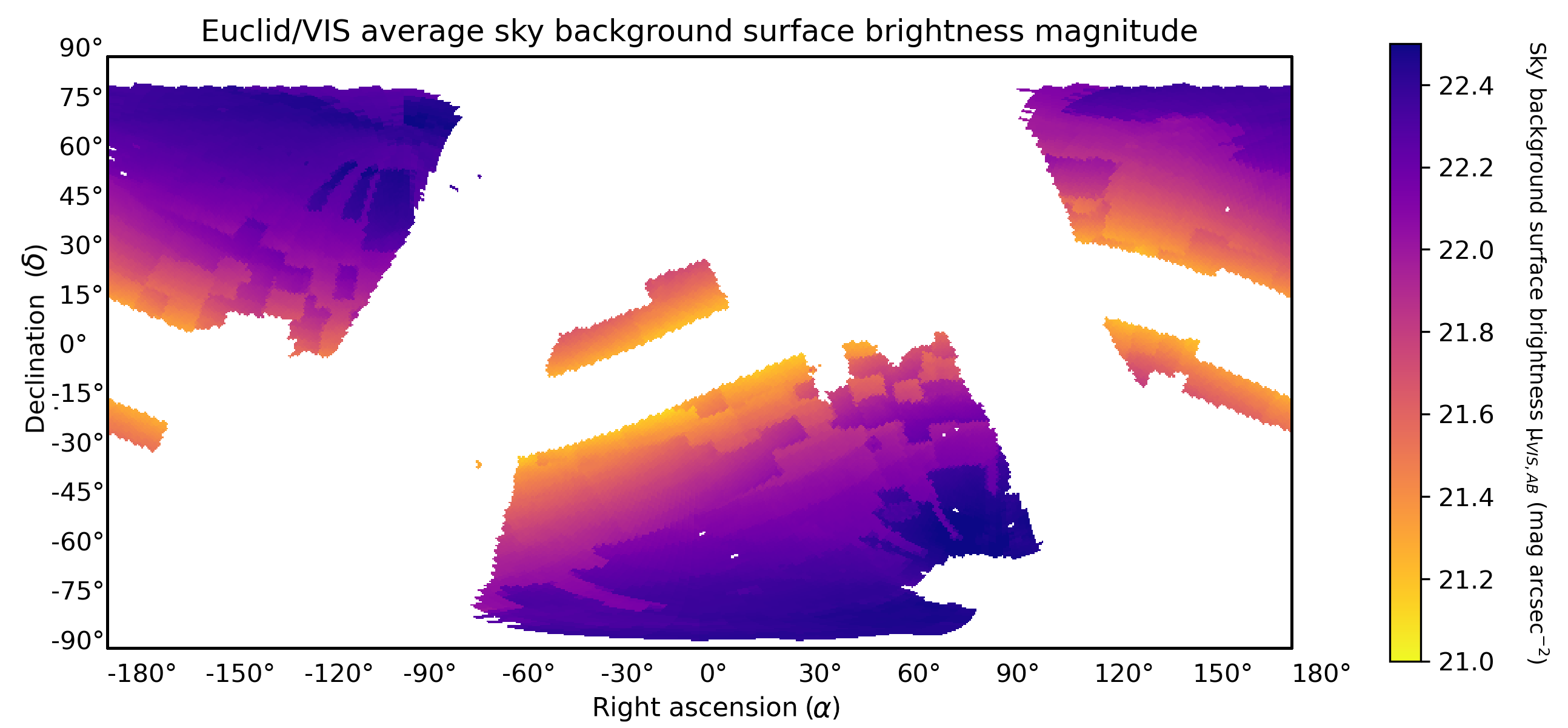}
\caption{Predicted sky background level map in equatorial coordinates for the \emph{Euclid}/VIS Survey \citep{Scaramella2021}, as a combination of zodiacal light, Milky Way interstellar medium (ISM), and the cosmic infrared background (CIB), using the NASA/IPAC sky background model. The resolution of the map is approximately $\ang{0.92;;}$. The sky background value of each bin corresponds to the mean value of the exposures contained inside. See the colorbar in the right.}
\label{fig:healpix_skysb}
\end{center}
\end{figure*}

Based on the methods discussed in the previous sections, the process to simulate the exposures can be summarized as follows:
\begin{enumerate}
    \item We select the pointings ($\alpha$, $\delta$, PA) in sequential order, starting from the first exposure for the four months to simulate (9916 exposures).

    \item We estimate the expected intensity level for the zodiacal light, ISM, and the CIB, following the methods described in Sect.\,\ref{Subsec:methods_skybackground} and \,\ref{Subsec:methods_straylight}. 
    
    \item We interpolate linearly the stray-light level at the different F1--F9 focal plane points in the simulated pointings (see Sect.\,\ref{Subsec:methods_straylight}) to obtain the predicted stray-light level and gradient for every independent pixel, including the components from the Solar System bodies, infield and outfield stars (see Fig.\,\ref{fig:straylight_results} in Sect.\,\ref{Subsec:results_stray}). By combining this with the previous step, we estimate the total sky background level (e$^-$) and its structure.
    
    \item Once the sky background components are combined, we simulate the effects of photon shot noise. First we transform the sky background array from electrons to photons dividing by the average QE. Then we generate an array of random Poisson values using the photon sky background as the $\lambda$ parameter as
    
\begin{equation}
\label{eq:poisson_noise}
    P(\lambda, k) = \frac{\lambda^{k} \mathrm{e}^{-\lambda}}{k!}\,.
\end{equation}

    \item We multiply the photon sky background image by the expected sensitivity non-uniformity of the camera (flat-field) at the simulation time (see Sect.\,\ref{Subsec:methods_flatfield}). 
    \item We transform the units of the array from photon to electrons. 
    
    \item We simulate contamination by CR's (see Sect.\,\ref{Subsec:methods_masking}). 
    
    \item We add dark current ($1.38\times10^{-6}$ e$^- $s$^{-1}$) and bias level ($9.6\times10^{3}$ e$^-$) according to the technical requirements.

    \item We simulate the effects of readout noise by adding Gaussian white noise with a standard deviation of $4.5$ e$^-$.   
    
    \item At this point, the simulated image closely resembles the properties of the expected raw images from VIS, with the notable exception that they lack any kind of astronomical source, apart from CR. We start the pre-calibration procedure by correcting the bias and dark current from the array.  
    
    \item We transform the units of the array from electrons to ADU ($3.5$ ADU per e$^-$). 
    
    \item Finally, we add the random pixel masks up to the filling factor described in Sect.\,\ref{Subsec:methods_masking} and Fig.\,\ref{fig:filling_fraction}.
    
\end{enumerate}

We show a completed exposure example in the right panel of Fig.\,\ref{fig:skysim_example}. The process described below is performed until we generate 9916 simulations, which correspond to approximately $120$ days of mission time (4 VIS exposures every 4252 seconds approximately), taking into account readout, dither, slew, and NISP observation time. Once the simulated observations have been generated, we normalize all the frames to their median value and carefully combine them together by using a bootstrapping median algorithm. 


\section{Results}
\label{Sec:results}

\begin{figure}[]
\begin{center}
\includegraphics[trim={0 20 25 45}, clip, width=0.49\textwidth]{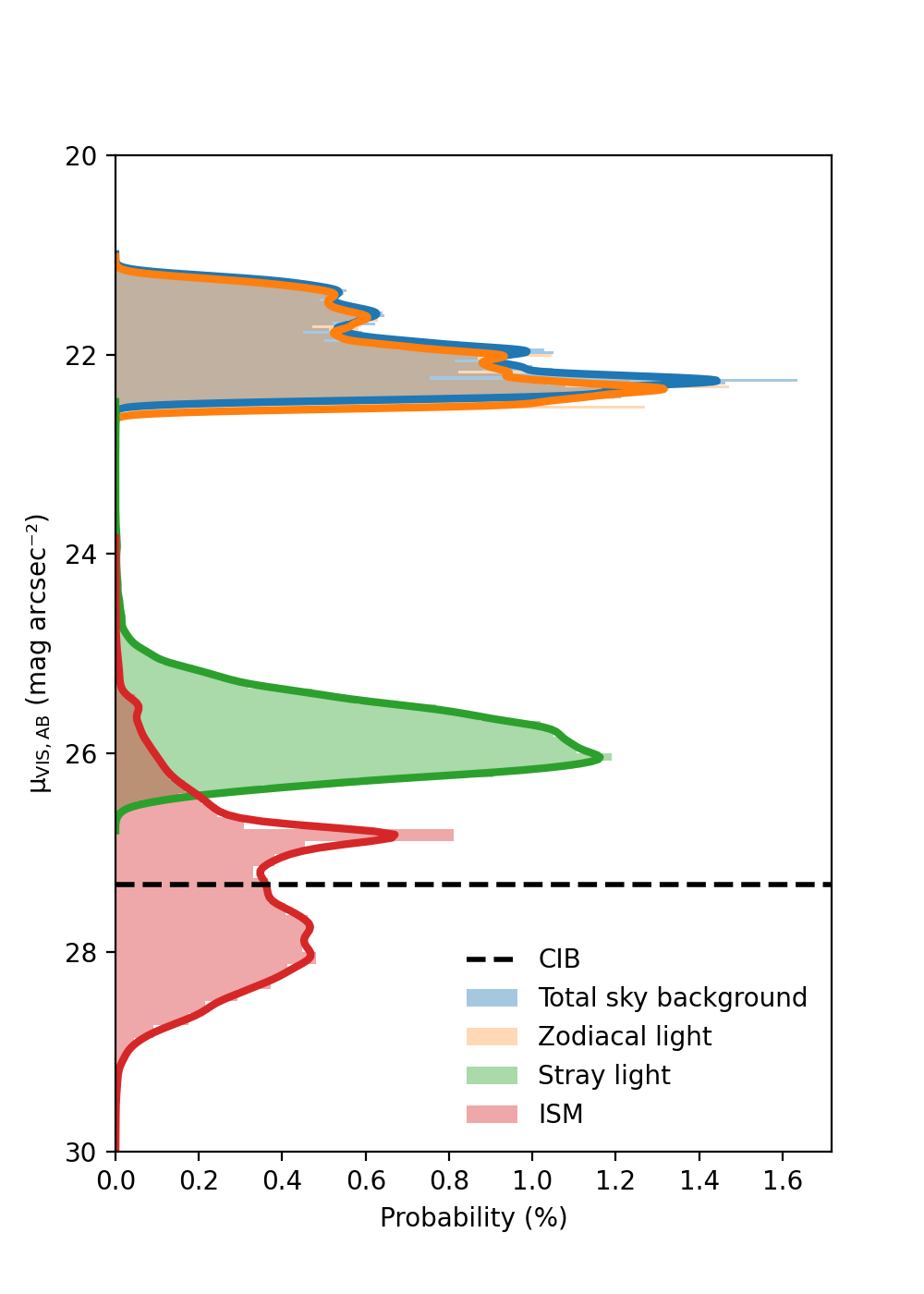}
\caption{Distribution of sky background surface brightness in the \emph{Euclid}/VIS exposures as predicted by the IRSA model for the zodiacal light (orange), interstellar medium (ISM, red), and the cosmic infrared background (CIB, black dashed line), the simulations for the average stray-light contamination (green), and the combination of all components (blue histogram).}
\label{fig:hist_VIS_surfmag}
\end{center}
\vspace{-0.5cm}
\end{figure}

In this section we summarize the results from this work. In Sect.\,\ref{Subsec:results_skybg} we study the surface brightness magnitude of the sky background for the different zodiacal light, ISM, and CIB components. Sect.\,\ref{Subsec:results_stray} details the analysis for the stray-light component. In Sect.\,\ref{Subsec:results_skyflat} we study the viability of the sky flat-field calibration strategy for the VIS detector, in terms of the field sensitivity correction precision and time resolution. Finally, in Sect.\,\ref{Subsec:results_limitingmagnitude} we provide a careful estimate of the expected limiting surface brightness magnitude for extended components that will be achievable for the survey.

\subsection{Sky background level}
\label{Subsec:results_skybg}

\begin{figure*}[t]
\begin{center}
\vspace{-0.75cm}
\includegraphics[trim={40 0 40 0}, clip, width=\textwidth, ]{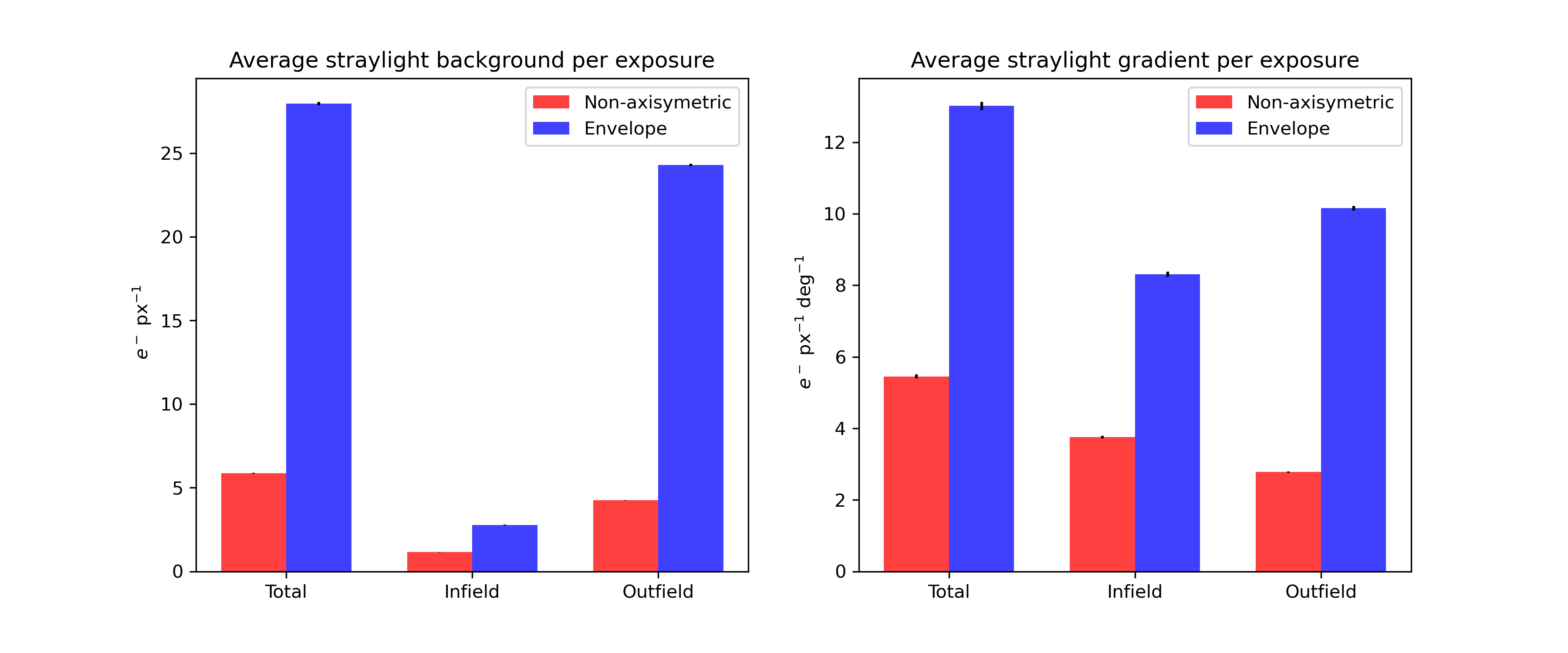}
\vspace{-0.75cm}
\caption{Stray-light contamination levels in the \emph{Euclid}/VIS Survey, taking into account sources inside (infield) or outside (outfield) the FOV, and the sum of all the objects (total). \emph{Left panel:} Average stray-light contamination level expected in the VIS exposures per pixel assuming the nominal 565s exposure time (e$^-$ px$^{-1}$). \emph{Right panel:} Average stray-light gradient level (e$^-$ px$^{-1}$ deg$^{-1}$) per exposure. \emph{Red bars:} Estimations based on the numerical non-axisymmetric NDI model. \emph{Blue bars:} Estimations based on the worst case scenario envelope NDI model. See the legend in the figures.}  
\label{fig:straylight_results}
\end{center}
\end{figure*}

Figure \ref{fig:healpix_skysb} represents the predicted map of surface brightness sky background for the \emph{Euclid}/VIS Survey, taking into account the zodiacal light, Milky Way ISM, and the CIB. For pure representation purposes, we do not include in this figure the scattered light introduced by stars, due to its high spatial variation. The average surface brightness of compact objects is not well defined for extended scales around $1\degree$ or larger, and their effect on the detectors have a dependence with the position angle (see Sect.\,\ref{Subsec:methods_straylight}). We found that the average surface brightness of the sky background ranges from $21.5$ \magarc\ at low latitudes to around $22.5$ \magarc\ at higher latitude regions. The regions associated with brighter sky background levels are dominated by the zodiacal light \citep[see][submitted]{Scaramella2014}. These parts of the survey are closer to the edges of \emph{Euclid} footprint towards the ecliptic plane, not only affected by the zodiacal light but also by the stray-light of the Solar System bodies (see Fig.\,\ref{fig:straylight_results}). 

The distribution of the surface brightness magnitude of the various background components is represented in the vertical histogram of Fig.\,\ref{fig:hist_VIS_surfmag}. The dominant component to the total sky background is the zodiacal light ($\muzody = 22.08^{+0.44}_{-0.78}$ \magarc). Based on the NDI model that takes into account the variation with the position angle, the second most important component is the stray-light from stars (we detail this result in Sect.\,\ref{Subsec:results_stray}). The rest of the components are much dimmer, with an average of $\muism = 27.5^{+1.3}_{-1.6}$ \magarc\ for the ISM. Nevertheless, dust cirrus can be much brighter, up to $\muism \sim 24$ \magarc, as observed in \citet{Mihos2017} and \citet{Roman2019}. The CIB appears as a constant background component of $\mucib=27.17$ \magarc. Therefore, the ISM background (i.e., Galactic cirri) is about 5 \magarc\ fainter than the zodiacal light background. Taking into account that the ISM structures are also different from exposure to exposure, they average out and are negligible in the computed sky flats. 



\subsection{Stray-light contamination}
\label{Subsec:results_stray}

 

Our results show that stray-light will generate an average surface brightness magnitude of $\mustray = 25.86^{+0.30}_{-0.37}$ \magarc\ in the VIS exposures, assuming the numerical NDI model (dependent on the position angle and the position in the FOV). Interestingly, if we assume the axisymmetric envelope model for the NDI (described in Eqs.\,\ref{eq:NDI_envelope}--\ref{eq:n_NDI_envelope}) the stray-light brightness estimation increases about $1.7$ \magarc, to $\mustray = 24.15^{+0.24}_{-0.27}$ \magarc. This discrepancy is anticipated: The axisymmetric NDI is a worst-case scenario that does not take into account the full baffling effects of the telescope optics, thus artificially increasing the contamination by nearby stars in the FOV if we compare it with the more realistic non-axisymmetric model. 

In Fig.\,\ref{fig:straylight_results} we present a summary of the results on the stray-light analysis, depending on the assumed model. In the left panel, we present the absolute stray-light flux as a function of the NDI model and the source. Estimation of the stray-light gradients is presented in the right panel. We differentiate between the infield and outfield stray-light components. The results show three additional important results:

\begin{enumerate}
    
    \item Out-of-field sources are responsible for approximately $80\%$ of the total amount of stray-light (76.1\% according to the non-axisymmetric NDI model and 88.9\% according to the envelope NDI model). 
    
    \item Total intensity of the stray-light gradients produced by the infield and outfield sources is similar, but their value differ significantly depending on the NDI model used. The symmetric NDI model predicts gradients twice as large ($\Delta S = 13.02^{+0.05}_{-0.04}\,$ e$^-$ px$^{-1}$ deg$^{-1}$) as those estimated using the non-axisymmetric NDI model ($\Delta S = 5.43^{+0.02}_{-0.01}\,$ e$^-$ px$^{-1}$ deg$^{-1}$).
    
    \item Interestingly, we found that the sum of the stray-light gradients from infield and outfield sources differ from the stray-light gradients measured taking into account all sources. An explanation for this effect is that outfield sources create gradients with higher intensity towards the edges of the FOV, while infield sources should create the opposite effect. On average, infield and outfield gradients neutralize partially when summed. 

\end{enumerate}

From the zodiacal model we estimate that the average zodiacal induced gradient in the \Euclid/VIS exposures is $0.598\pm0.001$ e$^{-}$ deg$^{-1}$. This is approximately 10 times less intense than the expected stray-light gradients per exposure. As a reference, for a surface brightness level of 22.5 \magarc\ (corresponding to the darkest regions of the \Euclid/VIS footprint) we would expect to have a FOV corner-to-corner change of $\Delta\mu=0.005$ \magarc\ (or 0.073\% of the total light per CCD). We thus consider that compared to the stray-light gradients, the zodiacal light gradients are negligible for our estimations.  

However, in the sky-flats, if the directions of the gradients are approximately random, then they will be partially suppressed by coadding multiple exposures. Nevertheless, there are two different facts that may affect this hypothesis: 1) the non-axisymmetric design of the spacecraft sunshield of the \emph{Euclid} spacecraft and 2) preferential directions of the position angle of the exposures of the survey. Our simulations take into account all these effects by using the different NDI models and the real parameters of the survey plan ($\alpha$, $\delta$, position angle, epoch of each exposure start, relative positions of the Solar System bodies). 


We find that the median stray-light background varies within $2\,$e$^-$ across the focal plane (the largest difference between two focal plane points is $1.44^{+0.01}_{-0.03}\,$e$^-$ in the case of the non-axisymmetric NDI model and $2.17^{+0.10}_{-0.07}\,$e$^-$ in the envelope NDI model). The spatial distribution of the median stray-light strongly depends on the NDI model (see Fig.\,\ref{fig:flatness_NDI}, in Appendix A). The focal plane point F8 shows a significantly larger stray-light contamination than the rest of the focal plane, when the NDI envelope model is considered. In contrast, the non-axisymmetric envelope NDI model shows a more uniform distribution. As discussed previously, the most reasonable cause for model-dependence of stray-light uniformity is that the azimuthal variation of the NDI takes into account more accurately the baffling effect of the telescope. In the case of the NDI envelope, which is a worst-case axisymmetric function, the stray-light blocking effect is removed from the simulation. In that case, the stray-light level is higher, with an extreme increase of the contamination from out-of-field sources compared to the more complex non-axisymmetric model, and the stray-light gradients are higher (as observed in Fig.\,\ref{fig:straylight_results}). Based on these results and for the sake of completeness, the adoption of the more complete non-axisymmetric NDI model for our models is justified.



\subsection{Sky flat-fielding}
\label{Subsec:results_skyflat}

\begin{figure}[t]
 \begin{center}
\includegraphics[trim={0 0 0 0}, clip, width=0.5\textwidth]{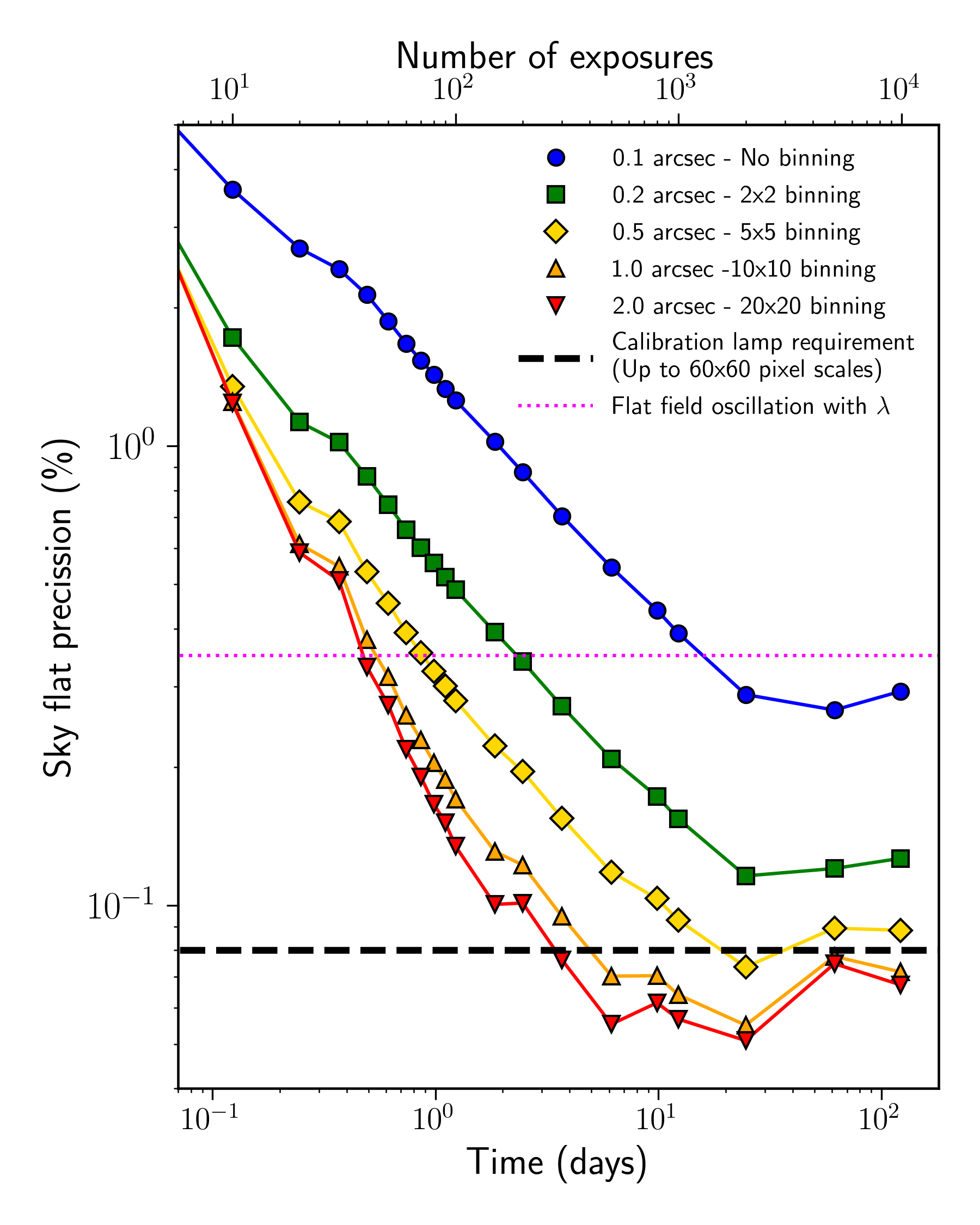}
\caption{Precision (\%) of the VIS sky flat-field as a function of the amount of combined exposures (as scaled by time) and the applied binning. The dotted horizontal magenta line represents the maximum measured variation of the LED flat-fields within the $\lambda$ range of VIS \citep{Szafraniec2016}. The black dashed horizontal line represents the precision requirement for the calibration lamp flat-fielding correction. See the legend for details.}
\label{fig:skyflat_SNR}
\end{center}
\end{figure}
\vspace{-0cm}

In this section we summarize the results of the sky flat-fielding simulations described in Sect.\,\ref{Subsec:methods_VIS_exposure_simulator}. In Fig.\,\ref{fig:skyflat_SNR} we show the efficiency of the sky flat-field correction as a function of the number of combined exposures. The precision of the sky flat-fields is measured as the $1\sigma$ width (defined as the $84.1-15.9$ percentile) of the probability distribution of the residuals between the sky flat-field estimation and the simulated PRNU of the detector (this is, the "true" flat-field), interpolated at the average epoch of the all exposures combined during that period.

The results shown in Fig.\,\ref{fig:skyflat_SNR} reveal a complex calibration scenario. In theory, coadding more images provide a better approximation of the flat-field, but in practice, degradation and stability effects over long periods of time actually increase the background residuals. The residuals of the sky flat-fields decrease rapidly following a near square-root power law as a function of the number of images combined during the first 10 days. Interestingly, our simulations show a strong deviation from the power law beyond that time ($> 500-1000$ images), where coadding more images to the sky flats does not help to reduce the background residuals in our images. This effect can be explained as a consequence of slight changes in the CCD detector sensitivity with time, which limit the integration time-span that we can use to generate the sky flats. Beyond a certain time period, changes in sensitivity are too high to be averaged in the independent images. Therefore, the adopted timescale for observing a set of images to generate sky flat-fields must be optimized in a trade-off between obtaining more SNR and avoiding the effect of the degradation of the effective throughput. As a consequence of this trade-off, the SNR of the sky flats cannot be improved beyond $\sim0.2\%$ in a pixel-to-pixel scale without applying some form of smoothing or spatial binning.


Nevertheless, recovering the pixel-scale structure in the sky flats is not a requirement for our purposes. The main objective of the sky flat-fields is to use them to correct the sensitivity at large spatial scales (using the CU lamp flats for the small scales), resulting in a high-SNR flat-field at considerably smaller spatial scales. Different types of sensitivity corrections can be used to obtain a valid calibration at all spatial scales (from pixel-to-pixel scales to the complete FOV). For that purpose, a viable strategy would be to first correct the exposures with the calibration lamp flat-fields and then co-add the resulting pre-calibrated images, obtaining a \emph{delta} sky flat. This technique has proven to be a valid method to correct large-scale gradients residuals in the flat-fields of the WFC3/IR and the ACS instruments of the \HST \citep{Pirzkal2011, ISR_Mack2017}. The flat-field calibration can thus be split into different components for large ($r>x$) and small ($r<x$) spatial scales: 


\begin{equation}
    R = F\,S + D\,t + B\, ,
    \label{eq:delta_sky_1}
\end{equation}
\begin{equation}
    R = (F_{r<x}\, F_{r>x})\, S + D\, t + B\, ,
    \label{eq:delta_sky_2}
\end{equation}
where $R$ is the raw image, $F$ is the flat-field, $S$ is the calibrated science image, $D$ the dark current per exposure time $t$ and $B$ the bias. A large-scale delta sky flat-field ($F_{r>x}$) can be generated after correcting the images with a first order flat-field ($F_{r<x}$) having spatial frequencies smaller than a certain scale $x$. The precision of the first order flat-field (calibration lamp) allows us to increase the SNR of the delta sky flats by smoothing or binning up to a certain scale. 

In Fig.\,\ref{fig:skyflat_SNR} we simulate what would be the precision obtained by using the delta sky flat correction at different binning scales (from $0.2$ to $2$ arcsec). Thanks to the high spatial resolution of \emph{Euclid}/VIS, with a minimal binning ($10\times10$ pixels$^2$, $1\times1$ arcsec$^2$) we will be able to meet the flat-field precision requirement every 5--10 days of the mission (in 3 days if the binning is made in $2\times2$ arcsec$^2$ scales). We must note that the calibration lamps will provide high-precision flat-fields for scales up to $60\times60$ pixels$^2$ ($6\times6$ arcsec$^2$), allowing us for a continuous correction of spatial sensitivity variations. 

In conclusion: Our simulations show that sky flat-fields can be periodically generated for scales larger than $>1$--$2$ arcsec combining the VIS science images obtained in periods of 3--10 days, complementing the calibration obtained using the on-board lamps. This result takes into account the technical specifications of the \emph{Euclid} spacecraft, the VIS instrument and its survey (i.e, sensitivity, exposure time, attitude, instrumental noise), as well as the observational strategy and the characteristics of the regions of the sky to be observed (zodiacal light, stray-light contamination, background source masking, cosmic-rays, ISM, CIB). Delta sky flat-fields generated using this method will be able to successfully complement the standard calibration procedure, providing a high-quality correction for large-scale sensitivity residuals, enabling the \emph{Euclid}/VIS Survey for the detection of large-scale low surface brightness structures. These sky-flats will be combined with self-calibration methods \citep{Manfroid1995} to correct for the largest spatial scales. We propose that a \emph{calibration ladder} (lamp flats for the small scales, self-calibration, and finally sky flat-fields) will enhance the precision of \emph{Euclid} to explore the low surface brightness Universe.


    

\subsection{\emph{Euclid}/VIS Survey surface limiting magnitude for extended sources}
\label{Subsec:results_limitingmagnitude}


One of the most important objectives of the present work is to provide a realistic prospect of the limiting surface brightness for extended sources in the \emph{Euclid}/VIS Survey. Taking advantage of our simulated frames, we can estimate the effect of a large variety of systematic errors in the actual limiting surface brightness. We define this limit as the corresponding surface brightness of a $3\sigma$ (percentile interval 0.13--99.86\%) intensity fluctuation measured over an area of $10\times 10$ arcsec$^2$, following the metric used in previous studies \citep{Trujillo2016, IACStripe82, Laine2018, Borlaff2019}. Note that this definition is arbitrary, and it is typically set to match the spatial scales of the low surface brightness features of nearby galaxies, which extend over larger sizes than one single pixel \citep{Mihos2017, Mihos2019}. 





\begin{figure*}[th!]
 \begin{center}
\includegraphics[trim={0 0 0 0}, clip, width=\textwidth]{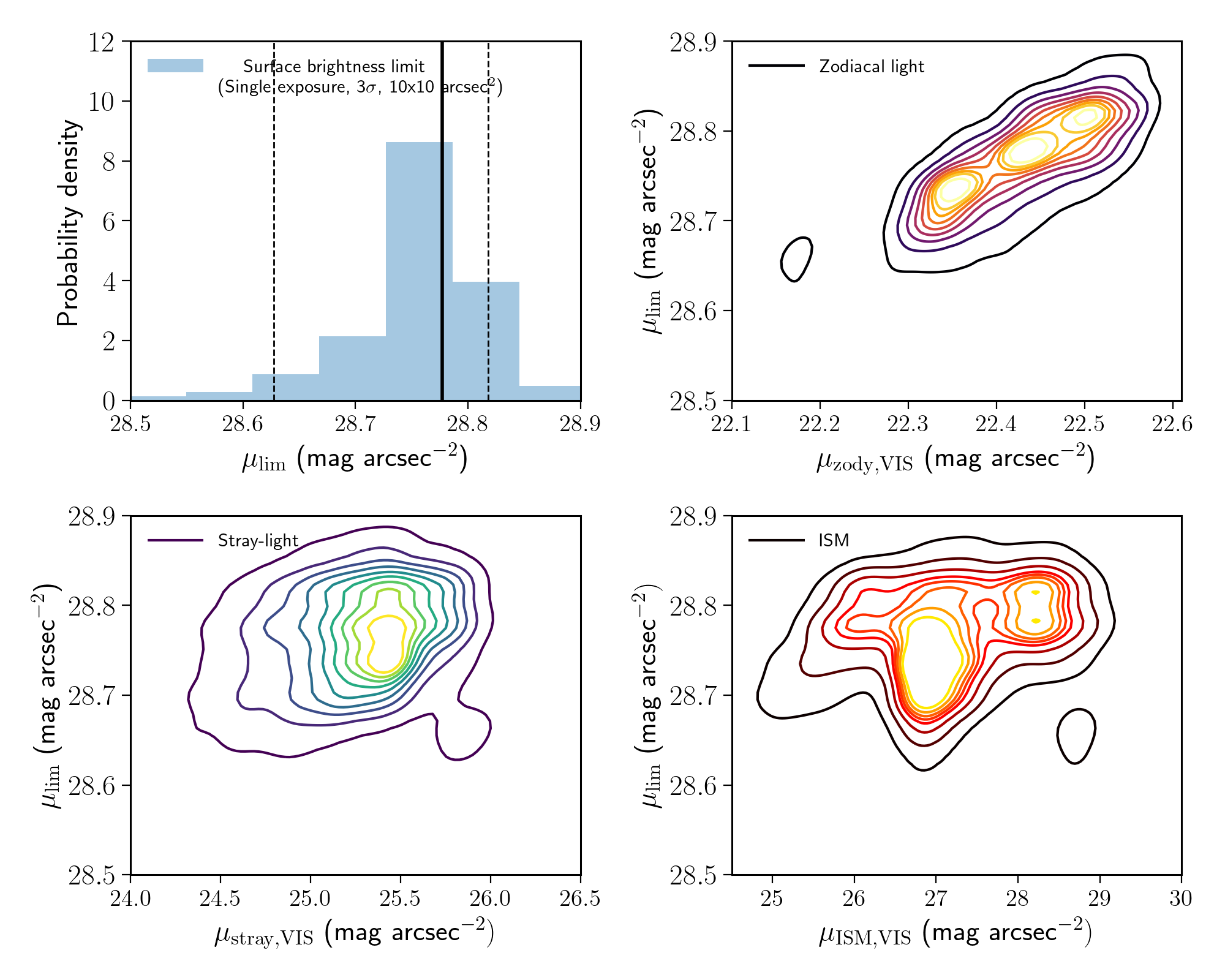}
\caption{Surface brightness limit of the individual simulated VIS exposures as a function of the different components of the sky background. \emph{Top left panel:} Histogram of surface brightness limit per exposure ($9916$ simulations). The solid and dashed vertical lines represent the median value of the distribution and its $\pm1\sigma$ dispersion percentiles. \emph{Top right panel:} Zodiacal light surface brightness vs. limiting surface brightness. \emph{Bottom left panel:} Average stray-light light level (from Solar System bodies and stars) surface brightness vs. limiting surface brightness. \emph{Bottom right panel:} Interstellar medium surface brightness vs. limiting surface brightness. Contours represent nine equidistant levels of probability density between 10\% and 90\%.}
\label{fig:limiting_magnitude}
\end{center}
\end{figure*}

In Fig.\,\ref{fig:limiting_magnitude} we show the results of the surface brightness limit analysis as functions of the different sky background components. We estimate limit surface brightness of $\mulim = 28.78^{+0.08}_{-0.28}$ \magarc\ per exposure, with a standard exposure time of 565 s. This limit was measured based on the results from the simulated images, after including the effects of the background light (see Sects.\,\ref{Subsec:methods_skybackground} and \ref{Subsec:methods_straylight}), Poisson noise, count discretization, and readout noise (see Sect.\,\ref{Subsec:methods_VIS_exposure_simulator}). We found that there is a clear strong dependence of the surface brightness limit with the intensity of the zodiacal light, which clearly dominates over the rest of the components \citep[][submitted]{Laureijs2011, Scaramella2014}. This effect is specially notable in the case of some of the calibration fields, which present much higher zodiacal light levels and thus lower surface brightness limits \citep{Scaramella2021}. Secondary effects like stray-light and the intensity of the ISM (i.e., cirrus) do not present any correlation with the limiting magnitude, showing that they are not dominant factors for the surface brightness limit. This result confirms that the zodiacal light is the main limitation of the mission in terms of depth, over the respective contributions on the stray-light, ISM or the CIB. 

\begin{figure*}[]
 \begin{center}
\includegraphics[trim={60 80 80 100}, clip, width=\textwidth, ]{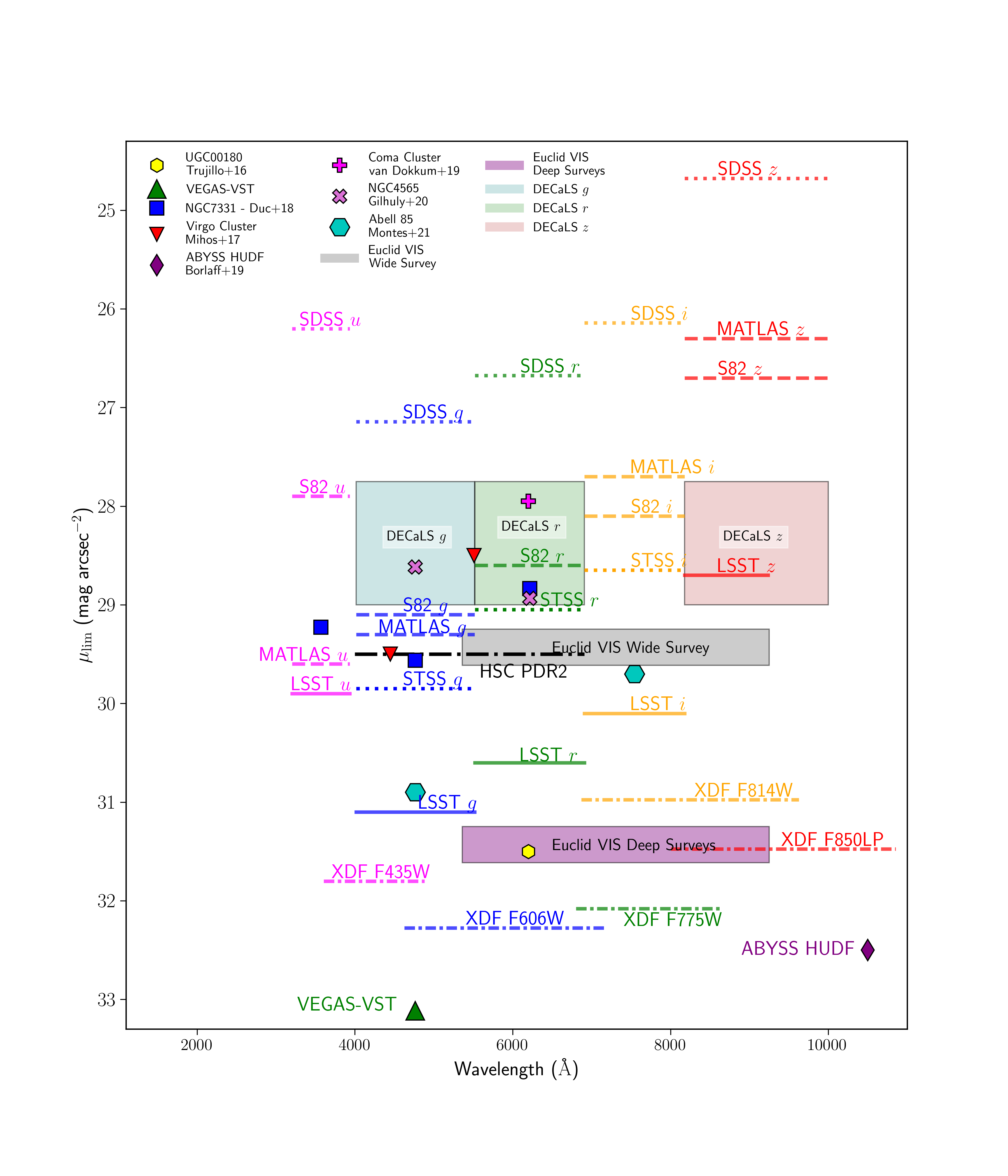}
\caption{Comparison of the surface brightness limit ($3\sigma$, $10\times10$ arcsec$^2$) for the \emph{Euclid}/VIS Wide Survey (and two magnitudes deeper for the Deep Fields), compared with a selection of deep optical and near-infrared surveys including SDSS \citep{York2000}, IAC Stripe 82 \citep[S82][]{IACStripe82}, the MATLAS deep imaging Survey \citep{Duc2015}, DECaLS \citep{Dey2019}, Stellar Tidal Stream Survey \citep[STSS]{MartinezDelgado2019}, Hyper Suprime-Cam DR2 \citep{Aihara2018}, Coma Cluster Dragonfly observations \citep{vanDokkum2020}, HST WFC3 ABYSS HUDF \citep{Borlaff2019}, XDF \citep{Illingworth2013}, UGC00180 10.4m GTC exploration \citep{Trujillo2016}, the Burrell Schmidt Deep Virgo Survey \citep{Mihos2017}, the VEGAS-VST \citep{Iodice2020,Ragusa2021} and LSST \citep[10 years full survey integration,][]{RubinTelescope}.}
\label{fig:depth_history}
\end{center}
\end{figure*}

The dithering pattern of the Wide Survey will ensure that almost every single position will be observed in three or four consecutive exposures of 565\,s, dithered using an S-pattern \citep[$\Delta x, \Delta y = $\ang{;;0}, \ang{;;0}; +\ang{;;50}, +\ang{;;100}; \ang{;;0}, +\ang{;;100}; +\ang{;;50}, +\ang{;;100},  ][]{Markovivc2017}. This observing strategy ensures that  about 40\% of a survey field will be imaged three times, and 40\% four times \citep{Scaramella2021}. Taking into account the results for the independent exposures (see Fig.\,\ref{fig:limiting_magnitude}), the limit surface brightness for the Wide Survey will be nearly $0.74$ \magarc\ deeper than the individual frames, reaching $\mulim=29.53
^{+0.08}_{-0.28}$ \magarc\ ($3\sigma, 10\times10$ arcsec$^{2}$). For those regions where the exposures overlap three times, the surface brightness magnitude limit will be $\mulim=29.37
^{+0.08}_{-0.28}$ \magarc. The depth achieved in the Wide Survey will be then comparable to that of the observations made by CFHT Megacam on NGC7331 \citep{Duc2018} or the S82 observations \citep[275 deg$^2$,][]{IACStripe82}, but for $15\,000$ deg$^2$ of the sky, with a better PSF, lower sky background, and a much higher spatial resolution (see Fig.\,\ref{fig:depth_history}). In addition to the Wide Field, three additional fields are especially interesting for the LSB science case, the North, South, and Fornax \emph{Euclid} Deep Fields\footnote{\Euclid Deep fields: https://www.cosmos.esa.int/web/euclid/euclid-survey}, which will combine a higher density of exposures, reaching surface brightness levels up to 2 magnitudes deeper than the Wide Survey. 

As a comparison, we show in Fig.\,\ref{fig:depth_history} the expected depth for the \Euclid Survey with some of the most notable results from the literature. The Deep Field mosaics have the potential to trace extended structures deeper than the expected surface brightness limit of the Vera C. Rubin (LSST) final mosaics, complementing a lower covered area (65 deg$^2$ between the three \Euclid\ Deep Fields and 15\,000  deg$^2$ vs. 18\,000 deg$^2$ for Rubin) with a higher-resolution and deeper limit in surface brightness in a similar wavelength range (although with notably less spectral resolution), being comparable to the depths in the ACS HUDF \citep[$0.003$ deg$^2$,][]{Illingworth2013}. Future missions like \emph{MESSIER} \citep{VallsGabaud2017} expect to reach much lower surface brightness magnitude levels closer to $\mulim=34$ \magarc\ in optical bands and  $\mulim=37$ \magarc\ in UV. Nevertheless, we must stress that these results are only an approximation to reality. Real detection limits are subject to many factors not covered in our simulations, including additional sources of stray-light contamination, sensitivity degradation, sky background over-subtraction during image processing, or changes in the observing plan. The results in this section should be interpreted as the optimal result to be obtained with a pipeline optimized for low surface brightness detections (see Sect.\,\ref{Sec:Discussion}).

\section{Discussion}
\label{Sec:Discussion}

In the present work confirm that \emph{Euclid}/VIS Survey enables unprecedented discovery space besides the core science. The combination of both a Deep and Wide Surveys offers a unique opportunity to study the low surface brightness Universe with the benefits of space-based resolution. The \emph{Euclid} Legacy Archive will provide high-quality imaging data up to depths and extensions not observed before. 

In general, limiting surface brightness magnitude depends with the size of the objects to be detected. Integrating over larger areas allows to increase the precision for the detection of diffuse objects. This applies also to surface brightness profiles. In example, (assuming no cosmological dimming) for local galaxies with an angular size of $D\sim1$ arcmin, an image depth of $\muvis=29.5$ \magarc\ (3$\sigma$ detection, measured over an area of $10\times10$ arcsec$
^2$) in the VIS Wide Survey, \textbf{an outermost radial bin spatial resolution of 5 arcsec, the area to be integrated would be between 450 to 850 arcsec$^2$}, depending on the inclination (\ang{45;;} to face-on) of the galaxy. As a consequence, the limit for the surface brightness profiles of these nearby galaxies would range from 30.2 to 30.5 \magarc\ (2--3 magnitudes deeper than current SDSS $r$ data, with 10 times higher spatial resolution). Based on SDSS observations, we estimate that there are approximately $24\,000$ galaxies outside the Local Group with diameters larger than 1 arcmin (measured as the Petrosian diameter in the SDSS $r$-band), which will be observed in the \emph{Euclid} Wide Survey. This means that on average we will find one of these extended objects in every pointing. This fact alone has the potential to move the extra-galactic structural analysis at ultra low surface brightness ($\mu \gtrsim 30$ \magarc) from individual explorations to the statistics domain. Simulation-based studies predict that a hypothetical survey with a limiting magnitude fainter than $29$ \magarc\ would detect up to 10 accretion features around Milky Way-type galaxies at distances greater than 30 kpc from the host \citep{Johnston2008}. In fact, volume-limited samples of nearby galaxies detect that almost 14\% of the galaxies present diffuse features compatible with minor merger events at a limiting magnitude of 28 \magarc\ \citep{Morales2018}. These results are compatible with those from \citet{Bilek2020}, and might suggest a conflict between the fractions predicted by cosmological models and the observations.  

The \emph{Euclid}/VIS mosaics will provide unbiased photometry of the structure of objects with smaller sizes (we can expect $3\times 10^{6}$ objects with $D_{\rm Petro,r}>1$ arcsec in the Wide Survey), paying special attention to avoid sky background over-subtraction and/or residual gradients in the North, South and Fornax Deep Surveys. Their potential depth and area will enable comprehensive investigations of the extended structure of vast numbers of galaxies at moderate redshift ($z\sim1$--2) reaching depths similar to the current observations available on the local Universe, overcoming the effect of cosmological dimming \citep{Tolman1930,Tolman1934}. As an example, two magnitudes deeper than the predictions for the \emph{Euclid}/VIS Wide Field for objects at $z=1.5$ is equivalent to a rest-frame observation at $\mulim = 27.8$ \magarc, comparable to the S82 observations in the Local Universe. This combination of depth, area and high spatial resolution will support studies of the evolution of the outskirts of galaxies across the most recent history of the Universe ($z=0$--1.5). 
Star count methods are very efficient to explore diffuse, local Universe structures, reaching far beyond integrated photometry where their stellar populations can be resolved \citep{Butler2004, McConnachie2009, Ibata2009}. The combination of deep, wide and space-based observations is ideal for these explorations, as the maximum distance where they are applicable is highly limited by the spatial resolution of the images \citep[e.g., 16 Mpc using the Hubble Space Telescope;][]{Zackrisson2012}. In addition, the study of the tip of the red giant branch \citep{Mouhcine2005} and globular cluster population \citep{Rejkuba2012} provides a precise independent distance estimation. These explorations require high-resolution observations, where space-telescopes have an advantageous position. These techniques are crucial for the study of the ultra-diffuse galaxies in the Local Universe, where \emph{Euclid} could provide a statistical sample that could ease the debates about the dark matter presence in these objects \citep[see][and references therein]{vanDokkum2018, Trujillo2019, Montes2020}. Moreover, higher-resolution and wider-area deep observations will reveal a great number of dwarf low surface brightness galaxies, which remain undetected beyond the local Universe.

Nevertheless, a significant number of challenges remain to be solved to ensure the quality of such mosaics, which are beyond the scope of this paper. Even though our results show that \emph{Euclid} is particularly well shielded against stray-light, gradients will still be observed in individual observations, necessitating their fitting and removal. Careful masking of the sources, including the extended wings buried in the background noise \citep{gnuastro, teeninga15_improv_detect, Dey2019} is one of the greatest challenges of low surface brightness imaging. Due to this "buried" emission getting absorbed into the sky background model, sky background over-subtraction is a common issue in many surveys. The consequential negative effects on scientific results extend far beyond the outer structure of extra-galactic sources. Co-addition of images with different background gradients increases the noise level of the final mosaics. Blind source-detection maps are more likely to lose small objects if they are in a highly over-subtracted region. Moreover, a certain fraction of the light in the sky background is caused by PSF effects \citep{Slater2009,Sandin2014,Sandin2015}, which smear the signal from the brightest pixels to the surrounding regions of the detector. While PSF deconvolution methods yield reconstruction of the original distribution of light \citep{Trujillo2016, Borlaff2017}, or even stellar source removal \citep{Roman2019}, such processing can be successfully achieved only if the sky background subtraction is not too aggressive in removing the spread light.

Ghosts created by secondary reflections add another layer of complexity to the PSF correction problem. Novel modeling and subtraction methods such as the one described in \citet{Matlas_scattered_light} for the CFHT MegaCam might be particularly useful to correct the individual frames of VIS before coadding. Nevertheless, all the techniques described require the precise determination of the PSF at scale lengths of approximately twice the size of the structure to study \citep[see][]{Janowiecki2010, InfanteSainz2020}. Their effect was beyond the scope of the current paper, but we will study the impact of the PSF and ghosts for low surface brightness science with \emph{Euclid} images in a forthcoming publication.  

Galactic cirri are one of the many extended low surface brightness structures that we expect to find in the \emph{Euclid} Survey. Their complex filamentary structure  \citep{Miville-Deschenes2016} mimics that of the extra-galactic tidal structures \citep{Cortese2010}, making them extremely hard to fit and separate, even counting with high-resolution far-infrared data \citep{Mihos2017}. Unfortunately, such maps are not available for most all-sky surveys, and due to its almost fractal-like structure, lower resolution (4--5 arcmin) alternatives such as IRAS \citep{Miville-Deschenes2005}, \Planck \citep{Planck2016} or WISE \citep{Miville-Deschenes2016} might not be enough to correct the high-resolution images of \emph{Euclid}/VIS. Multi-wavelength methods based on deep, high-resolution optical photometry \citep{Roman2019} may be the key to isolate the optical diffuse emission by the cirri, enabling the study of the Galactic and extra-galactic low surface structures by separate. Identification of low surface brightness large-scale cirri using multi-wavelength data in VIS is a possibility yet to be explored.  

An interesting problem yet to be studied is the effect of Charge Transfer Inefficiency \citep[CTI]{Israel2015} in the extended, low surface brightness structures. CTI contamination causes spurious image trailing that increases over time due to radiation damage. In HST/ACS, CTI became a notable problem due to the trailing effect of warm pixels in the dark frames. We will explore self-correction methods as those presented in \citet{ISR_Mack2017} in a future publication. While systematic effects like hot, bad, saturated pixels, diffraction
spikes, persistence effects, satellite trails and residual fringe patterns can be automatically detected and masked using convolutional neural networks (CNNs) on the individual exposures \citep{2019ASPC..523...99P} these methods can also be applied to detect merger signatures \citep{Ackermann2018} and other tidal features (\citealt{Walmsley2019}, Martinez-Delgado et al. in prep). However, mitigation of potential biases due to the lack of large training samples and contamination by foreground and background sources requires further refinement of these techniques.

Finally, there is the problem of wavelength variation of the flat-field \citep{Stubbs2006}. On-ground characterization studies using calibration lamps on the CCD273 VIS detectors have shown a small but significant variation of the flat-field with wavelength \citep{Szafraniec2016}. The amplitude of this wavelength variation ranges from 0.9\% at 5500$\AA$ to 0.6\% at 8500$\AA$. At longer wavelengths, a pattern of concentric rings starts to be visible. The origin of this pattern is suspected to be silicon resistivity variations during the manufacturing process of the crystal. Interestingly, this wavelength-dependency is another aspect where methods such as sky flat-fielding might be superior to the calibration lamps, at least for extended sources. In order to correct a wavelength dependence of the flat-field, ideally one would know {\it a priori} the SED of each pixel covering the sky. As shown in Sect.\,\ref{Subsec:results_skybg}, most of the detector area will be dominated by the zodiacal light whose SED may be matched using some combination flat-field generated with the on-board set of calibration lamps. Sky flat-fields perform this SED-dependent sky flat naturally by providing an estimate of the sensitivity independent on the calibration lamps, when constructed using the zodiacal light itself (the equivalent of a calibration lamp with the same SED as that of the observations). A combination of calibration lamp flat-fields for bright sources and sky flats for the dim regions may be the best solution for a successful calibration for all spatial and intensity ranges.

\section{Conclusions}
\label{Sec:Conclusions}

In the present paper we have studied the capabilities of the \emph{Euclid} space telescope as a low surface brightness observatory. Despite that the detection of dim extended sources is beyond the original nominal mission design, the characteristics of the telescope in terms of FOV, survey footprint, exposure time, sensitivity, wavelength coverage are ideal for this purpose. Nevertheless, systematic errors are often a major limitation for the study of the extended structure of dim objects and caused by 1) flat-field inaccuracy, and 2) stray-light residuals, which are extremely hard to predict and quantify.

Despite the fact that sky flat-fielding techniques have been proven to be successful in calibrating large-scale residual variations of the sensitivity, most of their application extend to ground-based observations, or NIR space observatories, where the sky background contribution is sufficiently bright for these calibrations \citep[with a few exceptions, see][]{ISR_Mack2017}. In addition, the asymmetric design of the \Euclid spacecraft external baffle could bias the sky flats. If that was the case, the position angles of the survey fields would need to be constrained for \Euclid\ legacy science. In this paper, we show that these effects are negligible.

In the present work we have studied the possibility of a low surface brightness reduction for \Euclid/VIS, taking advantage of the imaging data of the mission as an additional Legacy Science product. A key product of this investigation includes the development of a set of simulated background observations that takes into account the effects of all-sky stray-light contamination, zodiacal light, ISM, CIB, QE and payload transmission, instrumental and photon noise, cosmic-rays, flat-fielding and detector degradation. The results show that:

\begin{enumerate}

    \item The Wide Survey VIS mosaics have the potential to achieve a limiting surface brightness magnitude of $29.5$ \magarc\ in an area of 15\,000 deg$^2$.
    
    \item Sky flat-fielding is a valid strategy for the calibration of the \Euclid/VIS Survey. The science exposures will allow us to independently generate a high-quality delta sky flat correction every the 3--10 days (with a minimal spatial rebinning of $1\times1$ arcsec$^2$), complying with the calibration quality requirements of the mission.
    
    \item Stray-light will be efficiently shielded at $\lesssim 26$ \magarc\ in most frames. Gradients due to stray-light will be extremely low, and its average contribution to the sky flats is negligible. We confirm that the zodiacal light will be the main contributor to the sky background \citep{Laureijs2011}, with a magnitude of $\muzody = 22.08^{+0.44}_{-0.78}$ \magarc. 
\end{enumerate}

In addition to these results, the methods described in Sect.\,\ref{Subsec:methods_straylight} provide a prediction of the shape of the stray-light background in the individual frames on a pixel-by-pixel basis. The methods presented in this work allow for individual corrections of the stray-light in the \emph{Euclid} images, resulting in a more precise determination of the sky background over the standard \Euclid\ processing pipeline.





The limiting surface brightness magnitude of the final \emph{Euclid} mosaics will depend on how all the instrumental systematic effects are corrected. Considering the properties of the mission, we estimate that \emph{Euclid}/VIS will provide high-resolution imaging with limiting surface brightness close to $\mulim\ =29.5$ \magarc\ in the Wide Survey and two magnitudes deeper in the Deep Surveys. \emph{Euclid}'s extraordinary combination of sensitivity, angular resolution and sky coverage will support multiple, transformative scientific investigations including: 1) the study of extended discs, satellites and stellar halos as tracers of the dark matter distribution on galaxies, 2) unprecedented mapping of the zodiacal light and Galactic dust cirri, and 3) precise measurement of the anisotropies of the CIB.

\emph{Euclid} has the potential to be the next breakthrough in the understanding of the formation and evolution of galaxies, providing high-resolution, deep, and extremely wide imaging of the low surface brightness Universe to the scientific community, becoming a cornerstone of low surface brightness astronomy for the next decades.

\begin{acknowledgements}
The authors thank Françoise Combes, Emmanuel Bertin, and Mischa Schirmer for the provided input that helped to improve this publication significantly. We thank Koryo Okumura for his help with the stray-light modeling and prediction methods. We also thank Matthieu Marseille, Ruyman Azzollini, Stefano Andreon, Henry Joy McCracken and Kenneth Ganga for their contributions and comments to this manuscript. We give special thanks to Jason Rhodes and Jean-Gabriel Cuby for their support. Without your insight and feedback this project would have never been possible to finish. A.B. also thank Michael Fanelli for his support and interesting comments on the project. A.B. was supported by an appointment to the NASA Postdoctoral Program at the NASA Ames Research Center, administered by Universities Space Research Association under contract with NASA, and the European Space Agency (ESA), through the European Space Astronomy Center Faculty. We acknowledge \AckInstitutions. This work has made use of data from the ESA mission {\it Gaia} (\url{https://www.cosmos.esa.int/gaia}), processed by the {\it Gaia} Data Processing and Analysis Consortium (DPAC, \url{https://www.cosmos.esa.int/web/gaia/dpac/consortium}). Funding for the DPAC has been provided by national institutions, in particular the institutions participating in the {\it Gaia} Multilateral Agreement. This research made use of NumPy \citep{van2011numpy}, Astropy, a community-developed core Python package for Astronomy \citep{2013A&A...558A..33A}. All of the figures on this publication were generated using Matplotlib \citep{Matplotlib}. This work was partly done using GNU Astronomy Utilities (Gnuastro, ascl.net/1801.009) version 0.11.22-dc86.
\end{acknowledgements}

\bibliographystyle{aa}
\bibliography{ESAEuclid.bib}{}


\normalsize
\pagebreak
\pagebreak
\begin{appendix}


\section{Limitations of stray-light approximation}
\label{Appendix:limitations_straylight}

In Sect.\,\ref{Subsec:methods_straylight} we describe the method to accurately estimate the stray-light that the VIS focal plane receives from the all-sky distribution of stars. We based our analysis using the \emph{Gaia} DR2 catalog, with the addition of a 260 objects from the bright end ($G<3$ mag) detected by \citet{Sahlmann2016}. To be able to simulate the stray-light from every single object in the sky in a reasonable computational time, we use a combination of full-resolution \emph{Gaia} catalogs for the closest objects to the FOV and HEALpix binning for the stars beyond a certain critical radius ($R_{\rm min}$). In this Appendix we test the uncertainties and limitations associated with this approximation, estimating them as a function of the relative size of both low and high-resolution catalogs. 

\begin{figure}[t]
 \begin{center}
\includegraphics[width=0.5\textwidth]{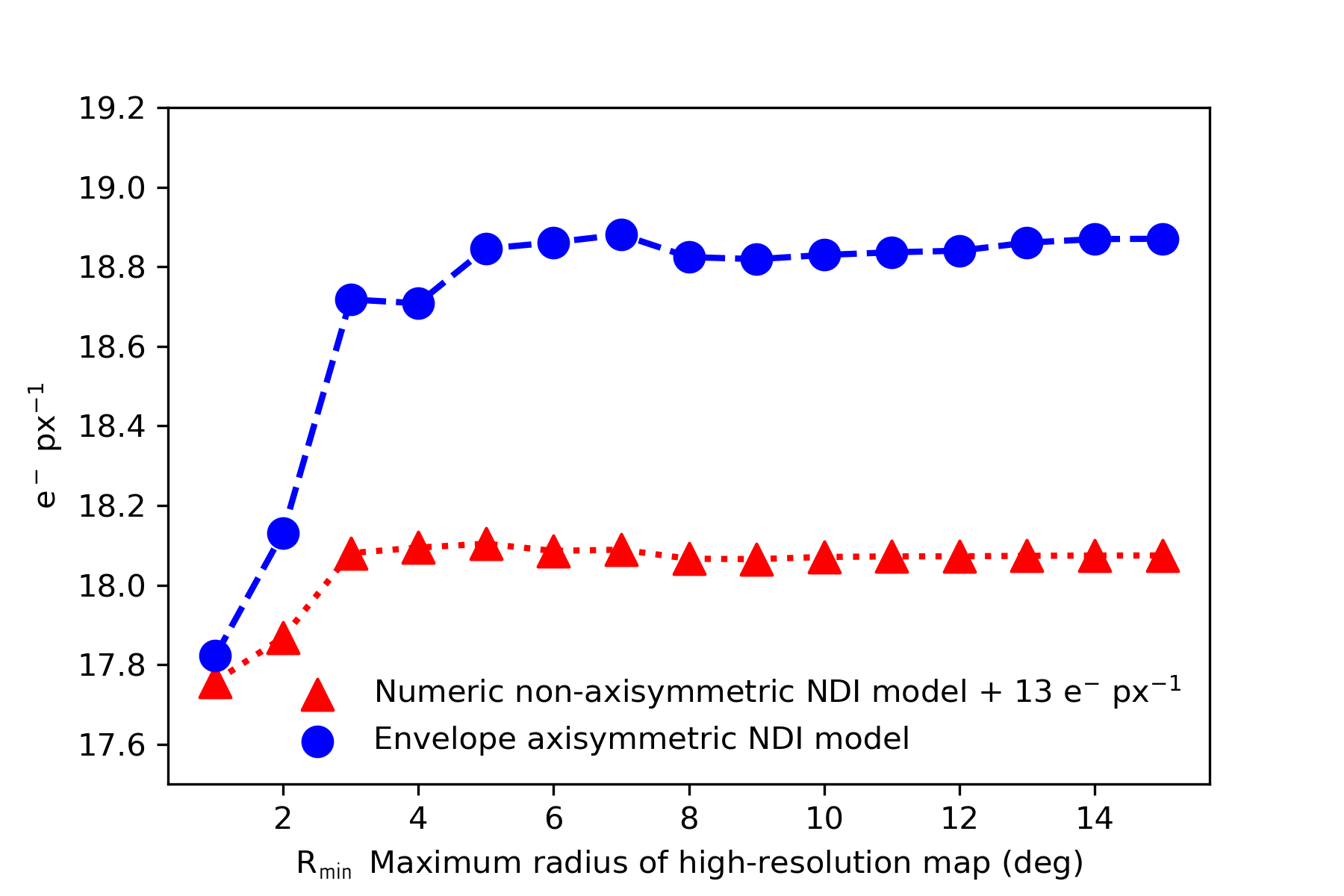}
\includegraphics[width=0.5\textwidth]{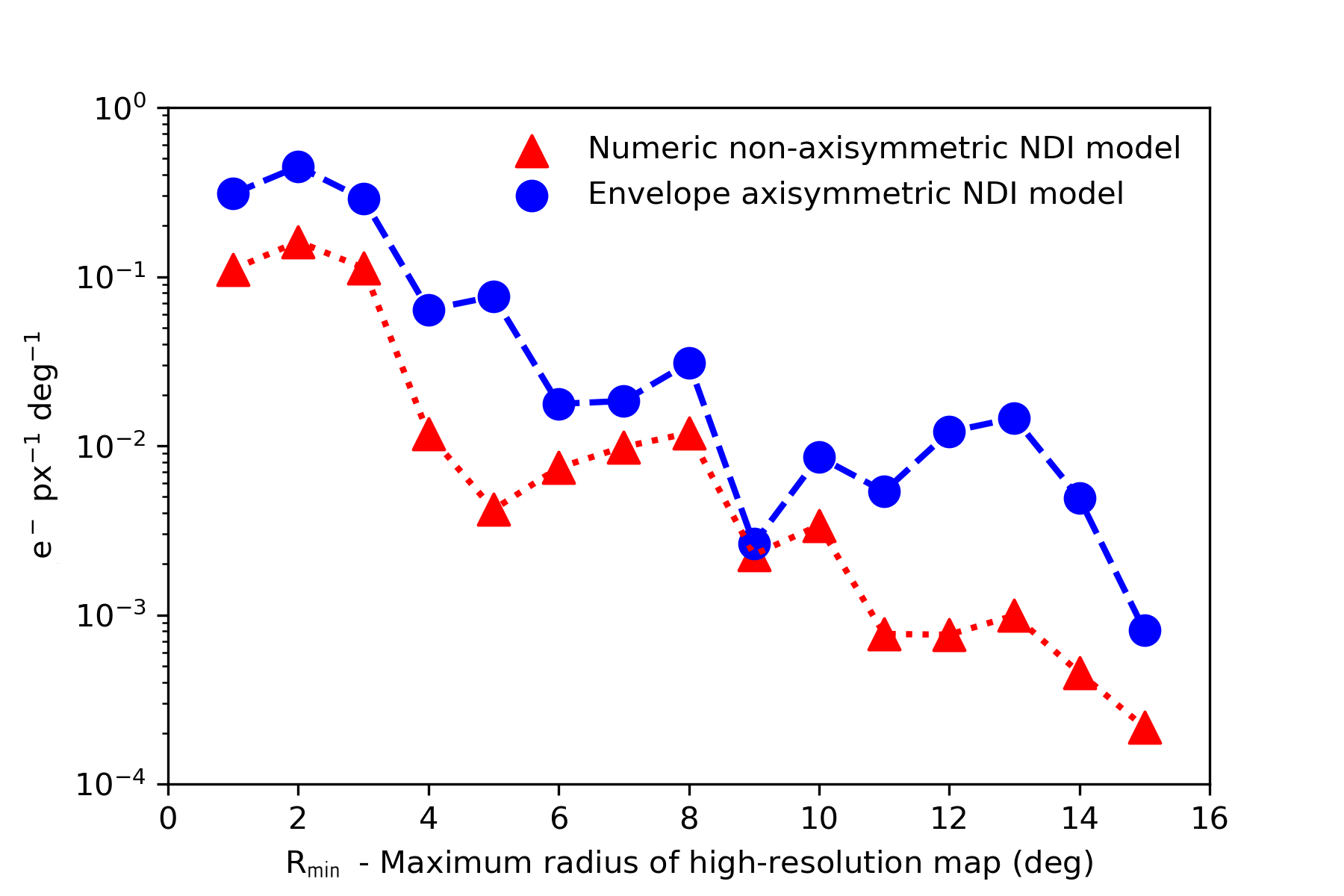}
\caption{Variation of the total stray-light estimated for a test sky pointing ($\alpha=\ang{-176.5;;}$, $\delta=\ang{+64.5;;}$, with a position angle of $\phi=\ang{25;;}$, with an increasing high-resolution radius $R_{\rm min}$ (see Sect\,\ref{Subsec:methods_straylight}), for both numerical non-axisymmetric, and envelope axisymmetric NDI models. \emph{Top panel}: Absolute flux of the estimated stray-light per pixel, in e$^{-}$ px$^{-1}$. \emph{Bottom panel:} Absolute variation of the estimated stray-light flux with increasing high-resolution radius. Notice how the the stray-light flux estimation converges to a semi-constant level for increasing $R_{\rm min}$.}
\label{fig:high_res_radius_straylight}
\end{center}
\end{figure}

In Fig.\,\ref{fig:high_res_radius_straylight} we show the dependence of the stray-light results with the variation of the maximum radius where we estimate the stray-light from each independent star from the catalog. When we consider a very small minimum radius, clustering all the objects in their HEALpix cells beyond $R > 1 \degree$, we obtain less stray-light than when considering a higher radius for the high-resolution map. Nevertheless, when considering higher values for $R_{\rm min}$ (simulating a higher region of the sky at full resolution) we find that the total stray-light level has a large increase for $R<2\degree$ but stays relatively constant beyond that limit. Moreover, analyzing the relative variation of the total stray-light with $R_{\rm min}$ (photons per pixel per degree increased in the high-resolution radius, see lower panel of Fig.\,\ref{fig:high_res_radius_straylight}) we find that the stray-light variation quickly converges to a constant value, finding differences lower than $0.1 $ e$^-$ px$^{-1}$ deg$^{-1}$ beyond $R > 5 \degree$. The reason for this behavior is that the NDI decreases very rapidly with radius, so slight changes in the position of the close stars make noticeable differences, but the effect of similar positional changes is negligible for stars at higher radii. We conclude that $R_{\rm min}=5 \degree$ is a safe limit to generate our simulations, as a compromise between computational effort and the precision of our simulations.

\begin{figure*}[t]
 \begin{center}
\includegraphics[trim={0 0 0 0}, clip, width=\textwidth, ]{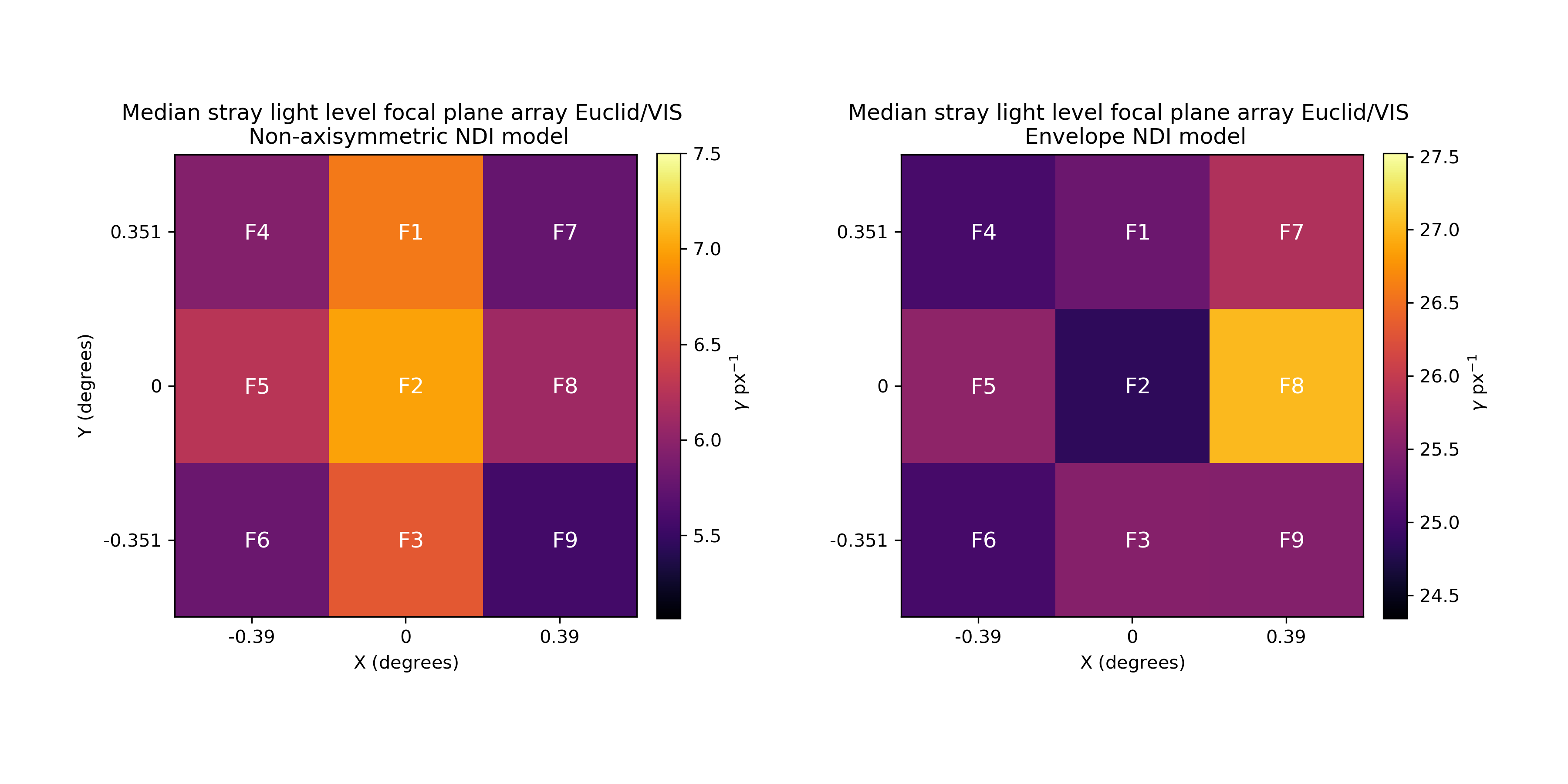}
\caption{Comparison of the median stray-light levels in the F1--F9 characteristic focal plane points of the \Euclid/VIS survey, taking into account sources inside (infield) and outside (outfield) the FOV for the non-axisymmetric NDI model (\emph{Left panel}) and the envelope worst-case NDI model (\emph{right panel}). See the colorbar in the figures.}  
\label{fig:flatness_NDI}
\end{center}
\end{figure*}

\end{appendix}
\end{document}